\documentclass[reqno,12pt,letterpaper]{article}
\pdfoutput=1
\usepackage{color, xcolor, colortbl}
\usepackage{graphicx,epstopdf}
\usepackage{geometry}
\usepackage{enumerate}
\usepackage{amsmath,amssymb,amsthm}
\usepackage{algorithm}
\usepackage{algorithmic}
\usepackage{caption}
\usepackage{comment}
\usepackage{subcaption}
\usepackage{bm}
\usepackage{appendix}
\usepackage{multirow}
\usepackage{braket}
\usepackage[english]{babel}
\usepackage{hyperref}
\usepackage[capitalize]{cleveref}
\usepackage[sort&compress,square,numbers]{natbib}
\usepackage{adjustbox}
\usepackage{xspace}
\usepackage[roman]{complexity}
\usepackage{soul}
\usepackage{tikz}

\usepackage{rotating}
\usepackage{setspace}
\usepackage{fancyhdr}

\usepackage{authblk}

\newtheorem{thm}{\protect\theoremname}
\theoremstyle{plain}
\newtheorem{lemma}[thm]{\protect\lemmaname}
\theoremstyle{plain}
\newtheorem{rem}[thm]{\protect\remarkname}
\theoremstyle{plain}
\theoremstyle{plain}

\theoremstyle{plain}

\providecommand{\definitionname}{Definition}
\providecommand{\assumptionname}{Assumption}
\providecommand{\corollaryname}{Corollary}
\providecommand{\lemmaname}{Lemma}
\providecommand{\propositionname}{Proposition}
\providecommand{\remarkname}{Remark}
\providecommand{\theoremname}{Theorem}

\newcommand\blfootnote[1]{%
  \begingroup
  \renewcommand\thefootnote{}\footnote{#1}%
  \addtocounter{footnote}{-1}%
  \endgroup
}

\newcommand{\norm}[1]{\left\lVert#1\right\rVert}

\newcommand{\REV}[1]{#1}

\title{On the Trotter Error in Many-body Quantum Dynamics with Coulomb Potentials}

\author[1,2]{Di Fang}
\author[3]{Xiaoxu Wu}
\author[4]{Avy Soffer}

\affil[1]{Department of Mathematics, Duke University, Durham, NC 27710, USA}
\affil[2]{Duke Quantum Center, Duke University, Durham, NC 27701, USA}
\affil[3]{Mathematical Sciences Institute, Australia National University, Canberra 2601, Australia}
\affil[4]{Department of Mathematics, Rutgers University, Piscataway, NJ 08854, USA}

\date{} 

\begin{document}
\maketitle

\begin{abstract}
Efficient simulation of many-body quantum systems is central to advances in physics, chemistry, and quantum computing, with a key question being whether the simulation cost scales polynomially with the system size. In this work, we analyze many-body quantum systems with Coulomb interactions, which are fundamental to electronic and molecular systems. We prove that Trotterization for such unbounded Hamiltonians achieves a $1/4$-order convergence rate, with explicit polynomial dependence on the number of particles.
The result holds for all initial wavefunctions in the domain of the Hamiltonian, and the $1/4$-order convergence rate is optimal, as previous \REV{work has numerically demonstrated} that it can be saturated by a specific initial ground state.
The main challenges arise from the many-body structure and the singular nature of the Coulomb potential. Our proof strategy differs from prior state-of-the-art Trotter analyses, addressing both difficulties in a unified framework. Our analysis treats the Coulomb potential as an unbounded operator without modification or regularization, and does not rely on spatial discretization, making it compatible with both first- and second-quantized circuit constructions.
\blfootnote{Emails:  di.fang@duke.edu; xiaoxu.wu@anu.edu.au; soffer@math.rutgers.edu.}
\end{abstract}

\tableofcontents

\section{Introduction}

Many-body quantum systems lie at the heart of a wide range of fundamental problems in physics, chemistry, and materials science. Despite their importance, simulating their dynamics remains a formidable challenge due to the exponential growth of the Hilbert space with the number of particles (system size). Quantum computing has emerged as a promising paradigm to tackle these problems, and many-body quantum dynamics simulation is widely regarded as one of its most compelling applications. Over the past decades, significant progress has been made toward simulating quantum systems with increasing complexity and realism.
A central question in this context is:

Can one demonstrate that a (quantum) algorithm can efficiently simulate many-body quantum systems with a cost that scales polynomially with the system size?

Addressing this question requires more than just analyzing the convergence order of the algorithm with respect to time steps or discretization parameters. Crucially, one must also carefully quantify the preconstants in the error bounds, particularly their dependence on the system size. For one-body or few-body problems, this dependence is often negligible or easily controlled, and standard analyses usually suffice. However, in the many-body regime, this dependence becomes highly nontrivial and constitutes a central aspect of both the error analysis and cost estimates, often necessitating the development of new theoretical understanding and techniques.

Significant progress and efforts have been made to understand the system-size dependence of quantum algorithms for various many-body settings, especially for finite-dimensional systems (e.g., spin systems with bounded Hamiltonians and second-quantized fermionic systems)~\cite{ChildsMaslovNamEtAl2018,AbramsLloyd1997,AspuruGuzikDutoiLoveHead-Gordon2005,BabbushBerryDominicKivlichanWeiLoveAspuru-Guzik2016,BabbushWiebeMcCleanMcClainNevenChan2018,KivlichanMcCleanWiebeGidneyAspuru-GuzikChanBabbush2018,StroeksLentermanTerhalHerasymenko2024}. Although the analysis becomes more difficult in the presence of unbounded operators, impressive advances have nevertheless been achieved in settings such as bosonic systems, quantum field theories~\cite{Preskill2019,TongAlbertMccleanPreskillSu2022,AbrahamsenTongBaoSuWiebe2023,PengSuClaudinoKowalskiLowRoetteler2023,SpagnoliWiebe2024,KuwaharaVuSaito2024,KuChenHuHsieh2025}, quantum harmonic oscillators~\cite{Somma2015}, and first-quantized systems with bounded or well-behaved potentials~\cite{KassalJordanLoveMohseniAspuru-Guzik2008,BabbushBerrySandersKivlichanSchererWeiLoveAspuru-Guzik2017,KivlichanWiebeBabbushEtAl2017,BabbushBerryMcCleanNeven2019,AnFangLin2021,AnFangLin2022,SuBerryWiebeEtAl2021,ChildsLengEtAl2022,RubinBerryKononovMaloneEtAl2023}. In many of these unbounded cases, a key technical ingredient is the use of case-dependent error analysis, which allows algorithmic cost estimates to be expressed in terms of the state norm concerning certain initial wavefunctions rather than worst-case error in terms of the operator norms~\cite{Somma2015,SahinogluSomma2020,AnFangLin2021,SuHuangCampbell2021,ZhaoZhouShawEtAk2021,ChildsLengEtAl2022,FangTres2023,BornsWeilFang2022,HuangTongFangSu2023,ZengSunJiangZhao2022,GongZhouLi2023,LowSuTongTran2023,ZhaoZhouChilds2024,YuXuZhao2024,ChenXuZhaoYuan2024,FangQu2025}.
This approach is particularly powerful in Trotterization-based algorithms, which play a unique role among Hamiltonian simulation methods. Unlike post-Trotter algorithms (such as Quantum Signal Processing, Quantum Singular Value Transformation, and truncated series methods)~\cite{LowChuang2017,GilyenSuLowEtAl2019,BerryChildsCleveEtAl2015,KieferovaSchererBerry2019,LowWiebe2019,BerryChildsSuEtAl2020,AnFangLin2022,FangLiuSarkar2025,BornsweilFangZhang2025} that incur operator norm dependence in their circuit implementation -- typically through the explicit or implicit use of block-encoding -- Trotterization avoids reintroducing such dependence. This makes Trotter formulas especially well-suited for simulating systems governed by unbounded Hamiltonians.

An important class of quantum systems is the many-body quantum systems with Coulomb interactions, which arise in fundamental applications such as electronic structure and molecular dynamics. Despite their significance, rigorous investigations of such systems remain relatively limited. The main challenge lies in the fact that both the kinetic and potential energy terms are unbounded operators, and the Coulomb potential is not only unbounded but also singular and non-smooth, violating the regularity conditions typically assumed in standard error analyses.

There have been a number recent advances in improving Trotter error estimates by incorporating the structure of the input state or observable~\cite{Somma2015,SahinogluSomma2020,AnFangLin2021,SuHuangCampbell2021,ZhaoZhouShawEtAk2021,ChildsLengEtAl2022,FangTres2023,BornsWeilFang2022,HuangTongFangSu2023,ZengSunJiangZhao2022,GongZhouLi2023,LowSuTongTran2023,ZhaoZhouChilds2024,YuXuZhao2024,ChenXuZhaoYuan2024,FangQu2025}. However, these results typically focus on many-body systems with bounded or regularized potentials, or finite-dimensional versions obtained through spatial discretization. In all such results, for the first-order Trotter formula, one gets the first-order convergence with respect to the number of Trotter steps. It is natural to assume that, as the number of spatial discretization degrees of freedom tends to infinity, the results would remain consistent with those for the underlying unbounded operator. However, recent findings have revealed a striking deviation: for systems with Coulomb interactions, the Trotter error can converge with only $1/4$-th order in the number of steps~\cite{BurgarthFacchiHahnJohnssonYuasa2024} for some initial wavefunction -- significantly slower than the first-order rate commonly expected. Theoretical analysis and \REV{physical} justification for this phenomenon has been provided in~\cite{BurgarthFacchiHahnJohnssonYuasa2024,BeckerGalkeSalzmannLuijk2024} for a specific eigenstate with a sharp 1/4 rate or in the one-body setting without a sharp rate. While the proof strategy can, in principle, be extended to many-body systems, it does not quantify how the error depends on the system size $N$, the critical factor for quantum algorithmic efficiency. This leads to an important open question:

\textit{Can we quantify the Trotter error for many-body quantum systems with Coulomb interactions, with explicit dependence on the system size?}

We answer this question affirmatively. In this work, we provide a rigorous error bound for Trotterization applied to many-body quantum systems with Coulomb potentials. Specifically, we prove that the Trotter error converges with order $1/4$ in the time step for any initial wavefunction in the domain of the Hamiltonian, with a polynomial dependence on the number of particles $N$. To the best of our knowledge, this is the first rigorous result of its kind. Our analysis opens the door to error and complexity estimates for simulating electronic and molecular systems, without smoothing or regularizing the singular Coulomb potentials, and paves the way for first-principle quantum simulations of such systems with provable efficiency. 

In this work, we focus on estimating the number of Trotter steps required for accurately simulating many-body quantum systems with Coulomb interactions. For recent progress on spatial discretizations in the first- and second- quantizations and significant advancements in quantum circuit design for such systems, we refer the reader to, e.g., ~\cite{SuBerryWiebeEtAl2021,BabbushHugginsBerryUngZhaoEtAlBaczewskiLee2023,RubinBerryKononovEtAlLeeNevenBabbushBaczewski2024,KuChenHuHsieh2025,BabbushWiebeMcCleanMcClainNevenChan2018,KivlichanMcCleanWiebeGidneyAspuru-GuzikChanBabbush2018,StroeksLentermanTerhalHerasymenko2024}. Our result well complements this significant line of research for providing the analysis for the Coulomb interaction as an unbounded operator. See \cref{sec:conclusion} for a discussion of practical implications and spatial discretization.

The rest of the paper is organized as follows. In \cref{sec:main_results}, we formally set up the problem and present our main results. We also outline the proof strategies and highlight key aspects of its novelty. \cref{sec:one-body} is devoted to the simple one-body case that serves to illustrate the core intuition behind our proof strategies. It also includes a short and elementary proof for the one-body case that already improves upon the best-known one-body estimates in the literature. \cref{sec:many-body_main} and \cref{sec:many-body-trotter} address the full many-body problem, corresponding to the two key steps in the one-body argument. In the many-body case, all preconstants must be explicitly quantified in terms of the system size -- unlike in the one-body setting, where they can be treated as fixed constants -- making the analysis substantially more delicate.

\section{Main Results and Proof Idea Overview}\label{sec:main_results}
In this section, we set up the problem and present our main results. We then discuss the proof strategy, highlighting key differences from prior state-of-the-art Trotter analysis approaches.

\subsection{Problem Setup and Main Results}

Given the system size (i.e. the particle number) $N \in \mathbb{N}^+$, we consider the Schr\"odinger equation with an $N$-body Coulomb potential:
\begin{equation}
    \begin{cases}
        i\partial_t \psi(t) = H\psi(t) \\
        \psi(0) = \psi_0 \in H^2\equiv H^2(\mathbb{R}^{3N})
    \end{cases}\qquad\qquad  t\in \mathbb R,\label{N-SE}
\end{equation}
where $ -\Delta := -\sum_{j=1}^N \Delta_{x_j},$ with $x_j \in \mathbb{R}^3$ for each $j = 1, \dots, N$ and $\psi(t)\equiv e^{-itH}\psi_0$. The interaction potential $V(x)$ is given by
\begin{equation}
    V(x) = \sum_{1 \leq j < k \leq N} \frac{c_{jk}}{|x_j - x_k|},\label{eq:N-V_def}
\end{equation}
where $c_{jk} \in \mathbb{R}, 1\leq j<k\leq N$, satisfies the uniform bound
\begin{equation}
    c_0 := \max_{1 \leq j < k \leq N} |c_{jk}| < \infty.\label{con: c0}
\end{equation}
We note that our formulation allows the Coulomb interactions that are attractive, repulsive, or a combination of both, thereby covering applications to both electronic and molecular systems.
In what follows, depending on the context, $\|\cdot\|$ denotes either the norm in $L^2 \equiv L^2(\mathbb{R}^n)$ of a wavefunction or the operator norm on $L^2(\mathbb{R}^n)$ of an operator. We also use $\|\cdot\|_{\mathcal{H} \to \mathcal{H}}$ to denote the operator norm on a Hilbert space $\mathcal{H}$, and $\|\cdot\|_{\mathcal{H}_{\REV{1}} \to \mathcal{H}_2}$ to denote the operator norm from a Hilbert space $\mathcal{H}_1$ to another Hilbert space $\mathcal{H}_2$. We adopt the following convention for the $H^2$ norm: for $g \in H^2$,
\begin{equation}\label{eq:def_H^2_norm}
    \|g\|_{H^2} := \sqrt{\|(-\Delta)g\|^2 + \|g\|^2},
\end{equation}
which is physically associated with the spatial curvature or high-frequency variations in the kinetic energy density.

For all our results, we consider initial conditions in the Sobolev space $H^2$, which is the domain of the unbounded Hamiltonian $H = -\Delta + V(x)$. This ensures that the Schr\"odinger operator makes sense when acting on the wavefunction -- i.e., the right-hand side of the Schr\"odinger equation, $H\psi$, is well-defined.

Our main result is summarized in the following theorem, which immediately implies that the number of Trotter steps required for quantum simulation scales only polynomially with the system size.
\begin{thm}[Long-time Trotter Error] \label{thm:main_trotter_long}
Let $H = A + B$ be the $N$-body Hamiltonian with Coulomb interactions given by~\cref{N-SE,eq:N-V_def,con: c0}, where $A = -\Delta$ denotes the kinetic part and $B = V(x)$ the Coulomb interaction potential.
Then for any initial state $\psi_0 \in H^2$, the long-time Trotter error for a total evolution time $T>0$ using $L$ time steps satisfies
\begin{equation}
    \norm{\left( e^{-iH T} - \left( e^{-iB T/L} e^{-iA T/L} \right)^L \right) \psi_0}\leq 
    \REV{\tilde{C} N^{4.5}} T t^\frac{1}{4} 
    \norm{\psi_0}_{H^2}
\end{equation}
where $t = T/L$ is the short-time Trotter step size, and \REV{$\tilde{C}>0$ is a universal constant depending only on $c_0$.}
\end{thm}
This $1/4$ rate of convergence with respect to the time step is optimal, as it was \REV{demonstrated} in~\cite{BurgarthFacchiHahnJohnssonYuasa2024} that there exists a specific initial eigenstate that achieves this rate both numerically and \REV{physically}. We emphasize that our result holds for all initial conditions in $H^2$, the domain of the Hamiltonian -- i.e. for any wavefunction for which the Schrödinger equation makes sense. However, this does not preclude the possibility that for certain specific initial conditions, the error may be significantly smaller. In other words, our estimate should be interpreted as a worst-case bound (analogous to the operator norm error bound in the finite-dimensional setting), rather than a specific-case analysis that focuses on initial states within a subspace of the Hamiltonian's domain. \REV{We do not attempt to optimize the universal constant $\tilde{C}$ in this work; a further discussion of its value appears in \cref{rem:universal_constant}.}

As a by-product of the proof, we also provide an estimate of the growth of the Sobolev norm in terms of system size $N$, which can be of independent interest.
\begin{thm}\label{thm:main2_H2_estimate} Under the same conditions of~\cref{thm:main_trotter_long}, the Sobolev norm of the solution $\psi(t)$ of~\cref{N-SE} at any time $t>0$ can be estimated as
\begin{equation}
    \|\psi(t)\|_{H^2}\leq C_{N}\|\psi_0\|_{H^2},
\end{equation}
with $C_N = \mathcal{O}(N^3)$ a constant depending only polynomially on the system size $N$ whose exact expression is given in~\cref{def CN} and independent of the time $t$.
\end{thm}

\subsection{Challenges and Proof Strategies}\label{sec:pf_strategy}
In this section, we outline the key challenges in analyzing many-body quantum systems with Coulomb interactions and highlight the novelty of our proof strategy.

It is well known that Trotterization exhibits a commutator scaling for bounded operators, see, e.g., 
~\cite{Trotter1959,ChildsSuTranEtAl2020,WiebeBerryHoyerSanders2010,ChildsSu2019,LowSuTongTran2023,SahinogluSomma2020,AnFangLin2021,SuHuangCampbell2021}. Specifically, for a Hamiltonian of the form $H = A + B$, where both $A$ and $B$ are bounded, the error between the first-order Trotter approximation $U_1(t) = e^{-iBt}e^{-iAt}$ and the exact evolution $U(t) = e^{-iHt}$ satisfies
\begin{equation}\label{eq:trotter_comm_A_B}
    \norm{U(t) - U_1(t)} \leq \frac{1}{2}\norm{[A,B]} t^2,
\end{equation}
per time step $t$. Since each step is unitary, the global error accumulates linearly in the number of steps, leading to first-order convergence in $t$ with an error constant proportional to the commutator norm $\norm{[A,B]}$.
However, this estimate breaks down immediately when the operators involved are unbounded, as in the case of Coulomb interactions. Even in the one-body setting, where $A = -\Delta$ and $B = 1/|x|$, both terms are unbounded on $L^2$ (we consider $L^2$ as it is the space for the wavefunctions and the unitary evolution preserves the $L^2$ norm), and the commutator 
\[
[\Delta, 1/|x|]
\]
is even more singular due to the nature of the Coulomb potential. 
Moreover, this commutator scaling as given in~\cref{eq:trotter_comm_A_B} should not hold in the unbounded case due to the breakdown of its derivation. In particular, it is derived from an exact error representation (see, e.g., \cite[Section 3.1]{ChildsSuTranEtAl2020}, \cite[Lemma 4]{AnFangLin2021}):
\begin{equation}\label{eq:err_rep_traditional_trotter}
U(t) - U_1 (t) = -\int_0^t d\tau \int_0^\tau ds \, e^{-i(t-\tau)H} e^{-is B}[B, A] e^{-i(\tau-s) B} e^{-i\tau A},
\end{equation}
which follows from a standard numerical analysis routine, such as using the variation-of-constants formula. In the finite-dimensional (bounded) setting, all operators map the same Hilbert space $\mathcal{H} \to \mathcal{H}$, allowing a straightforward norm bound on the right-hand side since all involved unitaries have operator norm one:
\begin{equation}
    \norm{e^{-iHt}}_{\mathcal{H} \to \mathcal{H}}
    =  \norm{e^{-iAt}}_{\mathcal{H} \to \mathcal{H}}
    =  \norm{e^{-iBt}}_{\mathcal{H} \to \mathcal{H}}
    = 1.
\end{equation}
For unbounded operators, however, this argument fails: the operator in~\cref{eq:err_rep_traditional_trotter} and the commutator $[A,B]$ are unbounded when considered as operators on $L^2$ (here $\mathcal{H} = L^2$ as all wavefunctions are $L^2$ normalized). Although the full Hamiltonian remains self-adjoint, it does not map $L^2$ to itself in the strong sense; rather, each operator makes sense (or acts) on its domain which is smaller than the whole Hilbert space $L^2$. For example, $-\Delta$ maps its domain $H^2$ (the Sobolev space; see the definition in~\cref{eq:def_H^2_norm}) to $L^2$. Consequently, the Trotter error must be analyzed in intermediate norm spaces, and care must be taken to track how these norms evolve under the dynamics.

Furthermore, while unitary operators preserve the $L^2$ norm, they do not, in general, preserve the norms of stronger spaces such as $H^2$:
\begin{equation}\label{eq:operator_norm_not_1}
    \norm{e^{-iHt}}_{L^2 \to L^2} =1, \quad \text{but  }
    \norm{e^{-iHt}}_{H^2 \to H^2} \neq 1.
\end{equation}
Physically, Sobolev norms are associated with kinetic energy and its higher-order structure. For instance, the $H^1$ norm corresponds to the kinetic energy (up to a constant), while the $H^2$ norm captures additional features related to the curvature or spatial variation of the wavefunction. In the presence of a potential, the kinetic energy -- and more generally, Sobolev norms -- are not conserved, although they typically remain uniformly bounded in time.
However, these bounds can depend sensitively on the number of particles and the interaction structure, making it essential to carefully quantify the system-size dependence in the analysis.

This marks one of the most significant distinctions between many-body analysis and both few-body and bounded-operator settings. In the case of bounded operators, the relevant unitaries have operator norm exactly one, requiring no further consideration. For unbounded operators in few-body systems, when system-size dependence is not tracked, the operator norms -- though not equal to one \REV{(see \cref{eq:operator_norm_not_1})} -- can be treated as fixed constants. In contrast, in the many-body setting, this simplification no longer holds: the relevant norms may scale with the number of particles, making it essential to explicitly quantify their dependence on system size. This introduces additional complexities into the analysis (see \cref{sec:many-body_main} for a detailed treatment of the norm estimate). 

Of course, the analysis involves more than just bounding the norms of the unitary operators. Every term in the error representation must be treated with care and in the correct order. In particular, \REV{the unitary generated by $-\Delta$ and that associated with the Coulomb potential act on different domains when viewed as operators mapping into $H^2$. Consequently, the sets of admissible wavefunctions that can be mapped into $H^2$ by these two evolutions are not the same.}
It turns out the ordering of the operators in the error representation also matters. Instead of the standard Trotter error representation as in~\cref{eq:err_rep_traditional_trotter},
we use the following alternative formulation
\begin{equation}\label{eq:our_trotter_error_rep_intro}
     U_1(t) - U(t) = i\int_0^t ds \, e^{-is B }[e^{-is A}, B]e^{-i(t-s)H},
\end{equation}
(see~\cref{lem:trotter_local_err_rep} for the proof). We have the unitary governed in $H$ on the right, and deliberately avoid further expanding the commutator 
$[e^{-is A}, B]$ into 
\begin{equation}\label{eq:expand_comm_eA_B}
  [e^{-is A}, B] =   -i \int_0^s \, d\tau \, e^{-i\tau A} [A,B] e^{-i(s-\tau)A},
\end{equation}
as the commutator $[A,B]$ for $A = -\Delta$ and $B$ as Coulomb interactions is even more singular compared to $[e^{-is A}, B]$. We note that having $e^{-itH}$ on the right is important. If instead we had a term involving $e^{-isB}$ on the right before the commutator -- as in \cref{eq:err_rep_traditional_trotter} -- then the commutator $[A, B]$ or $[e^{-isB}, A]$ would inevitably introduce some derivatives to the exponential $e^{-isB}$. Consider the one-body case as an example and, when $B = 1/|x|$, taking the first spatial derivative gives
\begin{equation}\label{eq:dx_of_exp_in_V}
\nabla \left( e^{-is/|x|} \right) = \frac{isx}{|x|^3} e^{-is/|x|},
\end{equation}
which is not in $L^2(\mathbb{R}^3)$ due to the singularity at the origin. We also note that although the two operator splitting orders -- taking $A = -\Delta$ and $B = V(x)$, or vice versa -- are mathematically equivalent, we choose $A = -\Delta$, $B = V(x)$ in our analysis, as it leads to expressions that are less singular and thus more amenable to control. To illustrate this at a high level, consider the one-body case as an example.
In the first case, where $A = -\Delta$ and $B = 1/|x|$, we have:
\begin{equation}
    [e^{-is\Delta}, 1/|x|] \psi(x) = \int_{\mathbb{R}^3} \left( \frac{1}{|y|} - \frac{1}{|x|} \right) K_s(x,y) \psi(y)\, dy,
\end{equation}
where $K_s(x,y)$ is the Schr\"odinger kernel. This presents a milder singularity moderated by the kernel. In contrast for the other order, the commutator $[e^{-is/|x|}, -\Delta] $ contains contributions like $\frac{1}{|x|^4}$, which is much more singular.

Using the exact error representation in~\cref{eq:our_trotter_error_rep_intro}, one key step is to identify a suitable intermediate Hilbert space  $\mathcal{H}_1$ such that 
\begin{equation}
    \norm{U_1(t)\psi_0 - U(t) \psi_0}_{L^2} \leq \int_0^t  \, ds \norm{e^{-isB}}_{L^2 \to L^2}\norm{[e^{-isA}, B]}_{\mathcal{H}_1 \to L^2}\norm{e^{-i(t-s)H} \psi_0}_{\mathcal{H}_1}, 
\end{equation}
for all $\psi_0$ in the domain of the Hamiltonian $H$. This provides a general strategy for deriving Trotter error estimates in the presence of unbounded operators. In the case of many-body Coulomb interactions, taking $\mathcal{H}_1 = H^2$ (the Sobolev space) suffices. This also highlights the importance of the order in the error representation: we prefer the unitary evolution governed by $H$ to appear on the right, as it maps any initial state in the domain of $H$ back into $H^2$. In contrast, if the rightmost unitary were governed by $V(x)$ alone, as explained in~\cref{eq:dx_of_exp_in_V}, it may fail to preserve the $H^2$ regularity due to the singularity in its derivatives.

In terms of the mathematical analysis, our proof technique is already novel and improves upon the state of the art even in the one-body setting. The previous state-of-the-art convergence rate for the one-body case~\cite{BeckerGalkeSalzmannLuijk2024} is $1/4-\varepsilon$ convergence for any $\varepsilon > 0$. That gap arose from the use of interpolation inequalities such as the Brezis–Mironescu inequality, which introduce unavoidable losses in the convergence rate. In contrast, we avoid such interpolation techniques entirely and instead use a cutoff method to handle the singular potential. Our result has a sharp $1/4$ convergence rate for the many-body case, and this rate is optimal, as it can be achieved by specific initial states both numerically and \REV{physically}~\cite{BurgarthFacchiHahnJohnssonYuasa2024}.

More on the cutoff method~\cite{SW1,SW2,SW3,SW4,SW5,SW-LD,breteaux2024light}: It is well known that the Coulomb potential $1/|x|$ belongs to $L^2 + L^\infty$ in $\mathbb{R}^3$: the singular part $1/|x|$ restricted to a unit ball is square-integrable, and its tail is bounded. Instead of directly splitting the potential based on this observation using the domain decomposition $|x| \leq 1$ or $|x|> 1$, we introduce a smooth cutoff decomposition depending on the time-step size:
\begin{equation}
    V(x) = V_\mathrm{reg}(x,s) + V_\mathrm{sin}(x,s),\qquad s\in (0,1],
\end{equation}
where the components $ V_\mathrm{reg}$ and $V_\mathrm{sin}$ are defined by
\begin{equation}\label{def: Vreg}
    V_\mathrm{reg}(x,s) := F\left( \frac{|x|}{s^\beta} > 1 \right) V(x)
\end{equation}
and
\begin{equation}\label{def: Vsin}
    V_\mathrm{sin}(x,s) := F\left( \frac{|x|}{s^\beta} \leq 1 \right) V(x)
\end{equation}
for a suitable $\beta \in (0,1)$, $F$ is a smooth cutoff function, and $s$ is related to the small Trotter step size. This allows us to isolate and control the singular behavior of the potential with greater precision. For the regular part, as it is essentially well-behaved as in the bounded operator setting, we can further use \cref{eq:expand_comm_eA_B}. For the singular part, we instead rely on a volume-based estimate. In the many-body setting, we treat each positional degree of freedom individually, using suitable changes of variables. For example, for a term of the form
\begin{equation}
    \frac{1}{|x_j - x_k|} = \frac{1}{|y|},
\end{equation}
we introduce the change of variables $y = x_j - x_k$, and define the cutoff function with respect to $y$. See~\cref{sec:many-body-trotter} for further details.

In the proof of the many-body setting, we establish the following lemma in \cref{sec:many-body_main} concerning many-body Coulomb potentials, which may be of independent interest. Note that this $N^{3/2}$ dependence is particularly appealing and unexpected, given that $V$ is a sum of $\mathcal{O}(N^2)$ terms.
\begin{lemma}\label{Key lem: N-body est_intro-ver} Let $V$ be the many-body Coulomb interactions as in~\cref{eq:N-V_def}, satisfying the condition~\eqref{con: c0}. Then, for all integer $N \geq 2$, we have
\begin{equation}\label{key est: N-body}
    \left\| V \frac{1}{|p|} \right\|_{L^2(\mathbb{R}^{3N}) \to L^2(\mathbb{R}^{3N})} \leq 2 c_0 N^{\frac{3}{2}},
\end{equation}
where $c_0$ is as defined in~\cref{con: c0} and the operator $\frac{1}{|p|}$ is defined according to the standard convention described in~\cref{eq:def_g(p)f}.
\end{lemma}

To make the presentation accessible, we illustrate the core ideas of this cutoff strategy in the one-body case in Section~\ref{sec:one-body}, and carry out the full analysis in the many-body setting in \cref{sec:many-body_main,sec:many-body-trotter}. The full many-body analysis requires significantly more delicate bookkeeping to track how various norms depend on the particle number $N$ and to ensure that all estimates remain polynomial in system size. In our proof (as laid out in \cref{sec:many-body_main,sec:many-body-trotter}), we treat the full many-body case.

\section{One-body Intuition}\label{sec:one-body}
In this section, we illustrate the intuition behind our proof using the Schr\"odinger equation with a one-body Coulomb potential:
\begin{equation}
    \begin{cases}
        i\partial_t \psi(x,t) = \left(-\Delta + \dfrac{c}{|x|}\right)\psi(x,t) \\
        \psi(x,0) = \psi_0 \in H^2(\mathbb{R}^3)
    \end{cases},
    \qquad\qquad t \in \mathbb{R}, \label{1-SE}
\end{equation}
where $-\Delta \equiv -\Delta_x$ is the Laplacian in $\mathbb{R}^3$, and $c \in \mathbb{R} \setminus \{0\}$ denotes a nonzero constant. Throughout the manuscript, we adopt the convention
\begin{equation}\label{eq:def_g(p)f}
g(p) f := g(-i\nabla_x) f,
\end{equation}
for any function $g$, which can also be interpreted in the Fourier sense.
The Fourier transform and its inverse are defined by
\[
\widehat{f}(\xi) := \frac{1}{(2\pi)^{n/2}} \int_{\mathbb{R}^n} e^{-i x \cdot \xi} f(x) \, dx,
\]
and
\[
f(x) := \frac{1}{(2\pi)^{n/2}} \int_{\mathbb{R}^n} e^{i x \cdot \xi} \widehat{f}(\xi) \, d\xi,
\]
for all \( f \in L^2(\mathbb{R}^n) \).

In the one-body setting, we use $C>0$ to denote a positive constant, which may vary from line to line. But in the many-body setting, we track the constants explicitly throughout the argument.

Let \( V(x) = \frac{c}{|x|} \) be the potential, and define the Hamiltonian of system (\cref{1-SE}) as \( H = -\Delta + V \). Let \( E(t) \) denote the error between the Trotterized evolution and the exact unitary dynamics (see~\cref{lem:trotter_local_err_rep}) for a short time interval $[0,t]$:
\[
E(t) = i \int_0^t ds\, e^{-isV} \left[ e^{-is(-\Delta)}, V \right] e^{-i(t-s)H}.
\]
\begin{thm}[One-body Short-time Trotter Error]\label{thm one-body}  
Let $0<t \leq 1$. There exists a constant \( C > 0 \) such that for all \( \psi_0 \in H^2 \), the following estimate holds:
\[
\| E(t)\psi_0 \| \leq C t^{\frac{5}{4}} \| \psi_0 \|_{H^2}.
\]
\end{thm}
This immediately implies the long-time error for the final time $T$ converges with a $1/4$ rate in the number of Trotter steps.
To prove \cref{thm one-body}, it suffices to establish two estimates stated in the following two lemmas, respectively. 
\begin{lemma}[Energy estimate -- Step 1]\label{energy lem: one-body} Let $H=-\Delta+\frac{c}{|x|}, c\in \mathbb R\setminus\{0\},$ be the Hamiltonian of system~\eqref{1-SE}. Then 
\begin{equation}
\sup_{t \in \mathbb{R}} \| e^{-itH} \psi_0 \|_{H^2} \leq C \| \psi_0 \|_{H^2}
\end{equation}
for some constant $C>0$.
\end{lemma}
\begin{lemma}[Commutator estimate -- Step 2]\label{com lem: one-body} Let $V(x)=\frac{c}{|x|}, c\in \mathbb R\setminus\{0\}$, be the potential of system~\eqref{1-SE}. Then
\begin{equation}
\| [e^{-is(-\Delta)}, V] \|_{H^2 \to L^2} \leq C s^{\frac{1}{4}}, \qquad \text{for all } s \in (0, 1],
\end{equation}
for some constant \( C > 0 \).
\end{lemma}

Step 1 essentially holds trivially in one- or few-body settings. This is because one can commute any power of $H$ with the Schr\"odinger equation~\cref{1-SE}, implying that the quantity $\norm{H^m \psi(t)}$ is preserved along the evolution. Uniform-in-time bounds then follow from the equivalence between this norm and the standard Sobolev norm. However, this approach relies on norm equivalence, which we avoid whenever possible, as the associated constants may introduce dependence on the system size in undesirable ways. To illustrate this point, consider a simple finite-dimensional example: let $x \in \mathbb{R}^{2^n}$. Although the $\ell^2$ and $\ell^\infty$ norms are equivalent, their ratio can grow polynomially with the Hilbert space dimension in the worst case:
\begin{equation}
    \norm{x}_\infty \leq \norm{x}_2 \leq 2^{n/2} \norm{x}_\infty.
\end{equation}
This is not a concern in one-body settings, but in the many-body regime, it becomes crucial to ensure that all constants depend only polynomially on the system size, which we carry out in detail in \cref{sec:many-body_main}. Instead of utilizing norm equivalence, we can consider the following route in the proof of Step 1, which is essentially what we use in the many-body setting. 

In the following, we sketch the proofs of both steps in the one-body case, with the full details presented in \cref{sec: Aux est one-body}. The proof of \cref{com lem: one-body} can be effectively regarded as the core of the one-body Trotter error analysis, as the remaining steps (such as Step 1) hold trivially. Nonetheless, we include a proof of \cref{energy lem: one-body} that avoids relying on norm equivalence. This approach is extended to the many-body case, allowing us to obtain explicit polynomial dependence on the system size.

\emph{Step 1: Energy estimate (idea of proving~\cref{energy lem: one-body}).} We start by using the identity
\begin{equation}\label{H id}
    (-\Delta)e^{-itH}\psi_0 = He^{-itH}\psi_0 - V e^{-itH}\psi_0 = e^{-itH}H\psi_0 - V e^{-itH}\psi_0,
\end{equation}
which, together with estimate~\eqref{Qineq}, reduces the problem to proving the following two bounds:
\begin{equation}\label{Hpsi0}
    \| H \psi_0 \| \leq C \| \psi_0 \|_{H^2},
\end{equation}
and
\begin{equation}\label{H1psi}
    \sup_{t \in \mathbb{R}} \| |p| e^{-itH} \psi_0 \| \leq C \| \psi_0 \|_{H^2}.
\end{equation}
To estimate the high-frequency part of $|p| e^{-itH} \psi_0$, we apply the identity~\eqref{H id} followed by estimate~\eqref{Qineq} once again. This shows that the bound in~\eqref{H1psi} can be reduced to proving~\eqref{Hpsi0}. Finally, estimate~\eqref{Hpsi0} follows directly from 
\begin{equation}
    \||p|f\|=\sqrt{(f,-\Delta f)_{L^2}}\leq \sqrt{\|-\Delta f\|^2+\|f\|^2}=\|f\|_{H^2}\qquad \forall\, f\in H^2 
\end{equation}
together with~\eqref{Qineq}:
\begin{equation}\label{H psi0}
    \| H \psi_0 \| \leq \| -\Delta \psi_0 \| + \left\| V \frac{1}{|p|} |p| \psi_0 \right\| \leq (1 + |c| C_{\mathrm{HLS},3}) \| \psi_0 \|_{H^2}.
\end{equation}

\emph{Step 2: Commutator estimate (idea of proving~\cref{com lem: one-body}).} The proof of \cref{com lem: one-body} is presented in \cref{pf:lem2_one-body}; here we present a sketch. To estimate the operator norm of the commutator 
\[
\left[ e^{-is(-\Delta)}, V \right]
\]
from \( H^2 \) to \( L^2 \), we decompose \( V \) into a smooth part \( V_{\mathrm{reg}} \) and a singular part \( V_{\mathrm{sin}} \), as defined in~\cref{def: Vreg,def: Vsin}. For \( V_{\mathrm{reg}} \), we establish the estimate
\[
\left\| [-\Delta, V_{\mathrm{reg}}(x,s)] f \right\| \leq \frac{C}{s^{\frac{3}{2}\beta}} \| f \|_{H^2},
\]
where the factor \( \frac{1}{s^{\frac{3}{2}\beta}} \) arises from the \( L^2 \)-norm of \( (-\Delta)V_{\mathrm{reg}}(x,s) \) in the \( x \)-variable (see also~\eqref{regV one-body} and~\eqref{est Vreg one-body}). This implies
\[
\left\| [e^{-is(-\Delta)}, V_{\mathrm{reg}}(x,s)] f \right\| \leq \int_0^s \frac{C}{s^{\frac{3}{2}\beta}} \| f \|_{H^2} \, du = C s^{1 - \frac{3}{2} \beta} \| f \|_{H^2}.
\]
For the singular part, we use the $ L^2$-norm decay (volume estimate) of $V_{\mathrm{sin}}(x,s)$:
\[
\left\| [e^{-is(-\Delta)}, V_{\mathrm{sin}}(x,s)] f \right\| \leq C \| V_{\mathrm{sin}}(x,s) \|  \| f \|_{H^2} \leq C s^{\frac{1}{2} \beta} \| f \|_{H^2}.
\]
Combining both estimates, we obtain
\[
\left\| [e^{-is(-\Delta)}, V] f \right\| \leq C \left( s^{1 - \frac{3}{2} \beta} + s^{\frac{1}{2} \beta} \right) \| f \|_{H^2} \leq C s^{\frac{1}{4}} \| f \|_{H^2},
\]
where we choose $\beta = \tfrac{1}{2}$ such that $1 - \tfrac{3}{2} \beta = \tfrac{1}{2} \beta$. This leads to the $1/4$ convergence rate.

\section{$N$-body Solution Norm Estimate}\label{sec:many-body_main}
In this section, we provide the proof of \cref{thm:main2_H2_estimate} for the many-body case (corresponding to Step 1 in the one-body intuition discussed in~\cref{sec:one-body}), namely, to prove
\begin{thm}\label{thmN} Assume that condition~\eqref{con: c0} holds. Then the solution \( \psi(t) \) to the system~\eqref{N-SE} lies in \( H^2 \) and satisfies the estimate
\begin{equation} \label{thmN est}
    \|(-\Delta)\psi(t)\| \leq \left(1 + 3 c_0 C_{\mathrm{HLS},3} N^{3/2} + 2 c_0^2 C_{\mathrm{HLS},3}^2 N^3\right) \| \psi_0 \|_{H^2},
\end{equation}
where $c_0$ and $C_{\mathrm{HLS},3}$ are some absolute constants (see \cref{con: c0,def CHLS3} for the explicit definitions). Moreover, we have
\begin{equation}
    \|\psi(t)\|_{H^2} \leq C_N \|\psi_0\|_{H^2},
\end{equation}
where \( C_N \) is defined by
\begin{equation} \label{def CN}
    C_N := 2 + 3 c_0 C_{\mathrm{HLS},3} N^{3/2} + 2 c_0^2 C_{\mathrm{HLS},3}^2 N^3.
\end{equation}
\end{thm}

We remark that our proof also yields the following estimate, which may be of independent interest:
\begin{equation}\label{eq:lap_norm_bounded_by_Hpsi}
    \norm{-\Delta \psi(t)} \leq 
    (1 + 2 c_0N^{3/2})\norm{H\psi_0} + \left(2 c_0N^{3/2}  + 4c_0^2 N^3\right)\norm{\psi_0},
\end{equation}
where we have substituted $C_{\mathrm{HLS},3}$ with its numerical value 2. The right-hand-side depends only on the initial wavefunction, and $\norm{H\psi_0}$ corresponds to the second moment of initial energy. This means that our main result,~\cref{thm:main_trotter_long}, can alternatively be expressed in terms of $\norm{H\psi_0}$ and $\norm{\psi_0}$. As both forms depend only on the initial state, we choose to present the $H^2$ version because it is simpler and more concise.

The proof of~\cref{thmN} relies on the following lemma, which provides an operator norm estimate for $V \frac{1}{|p|}$ from $L^2$ to $L^2$. This lemma, which is closely related to the Hardy–Littlewood–Sobolev inequality (see, e.g.,~\cite[Theorem 2.5]{Herbst1977}, \cite[(1.7)]{HLSineq}, \cite[Chapter V]{Stein1970}), will be proved at the end of this section after proving \cref{thmN}.
\begin{lemma}\label{Key lem: N-body est} Let \( V \) be as in~\cref{eq:N-V_def}, satisfying the condition~\eqref{con: c0}. Then, for all \( N \in \mathbb{N}^+ \setminus \{1\} \), we have
\begin{equation}
    \left\| V \frac{1}{|p|} \right\| \leq c_0 C_{\mathrm{HLS},3} N^{\frac{3}{2}},
\end{equation}
where \( c_0 \) and \( C_{\mathrm{HLS},3} \) are as defined in~\cref{con: c0,def CHLS3}, respectively.

\end{lemma}
\begin{proof}[Proof of~\cref{thmN}] Following the proof of~\cref{lem Hid}, but using~\cref{Key lem: N-body est} in place of~\eqref{Qineq}, we conclude that \( \psi(t) = e^{-itH} \psi_0 \in H^2 \) for all \( t \in \mathbb{R} \), provided \( \psi_0 \in H^2 \). Then the identity  
\begin{equation}\label{id: n-body}
    (-\Delta) e^{-itH} \psi_0 = e^{-itH} H \psi_0 - V e^{-itH} \psi_0
\end{equation}
holds. Applying~\cref{Key lem: N-body est} and using the unitarity of \( e^{-itH} \) on \( L^2 \), we obtain
\begin{equation}\label{line3}
\begin{aligned}
    \| (-\Delta) e^{-itH} \psi_0 \| 
    &\leq \| H \psi_0 \| + \| V e^{-itH} \psi_0 \| \\
    &\leq \| (-\Delta) \psi_0 \| + \| V |p|^{-1} \| \left( \| |p| \psi_0 \| + \| |p| e^{-itH} \psi_0 \| \right) \\
    &\leq \| (-\Delta) \psi_0 \| + c_0 C_{\mathrm{HLS},3} N^{3/2} \left( \| |p| \psi_0 \| + \| |p| e^{-itH} \psi_0 \| \right).
\end{aligned}
\end{equation}
Next, we estimate \( \| \chi(|p| > 1) |p| e^{-itH} \psi_0 \| \). Applying~\cref{id: n-body} to \( \chi(|p| > 1) |p| e^{-itH} \psi_0 \), we get
\begin{equation}
\begin{aligned}
    \chi(|p| > 1) |p| e^{-itH} \psi_0 
    &= \chi(|p| > 1) |p|^{-1} (-\Delta) e^{-itH} \psi_0 \\
    &= \chi(|p| > 1) |p|^{-1} e^{-itH} H \psi_0 
    - \chi(|p| > 1) |p|^{-1} V e^{-itH} \psi_0.
\end{aligned}
\end{equation}
By duality and~\cref{Key lem: N-body est}, we have
\begin{equation}
    \| |p|^{-1} V \| = \| V |p|^{-1} \| \leq c_0 C_{\mathrm{HLS},3} N^{3/2}.
\end{equation}
Using this estimate and the unitarity of \( e^{-itH} \), we obtain
\begin{equation}
\begin{aligned}
    \| \chi(|p| > 1) |p| e^{-itH} \psi_0 \| 
    &\leq \| \chi(|p| > 1) |p|^{-1} \| \cdot \| H \psi_0 \| 
    + \| |p|^{-1} V \| \cdot \| \psi_0 \| \\
    &\leq \| (-\Delta) \psi_0 \| 
    + \| V |p|^{-1} \| \left( \| |p| \psi_0 \| + \| \psi_0 \| \right) \\
    &\leq (1 + 2 c_0 C_{\mathrm{HLS},3} N^{3/2}) \| \psi_0 \|_{H^2}.
\end{aligned}
\end{equation}
Therefore,
\begin{equation}
\begin{aligned}
    \| |p| e^{-itH} \psi_0 \| 
    &\leq \| \chi(|p| \leq 1) |p| e^{-itH} \psi_0 \| 
    + \| \chi(|p| > 1) |p| e^{-itH} \psi_0 \| \\
    &\leq (2 + 2 c_0 C_{\mathrm{HLS},3} N^{3/2}) \| \psi_0 \|_{H^2}.
\end{aligned}
\end{equation}
Substituting this into~\eqref{line3}, we find
\begin{equation}
    \| (-\Delta) e^{-itH} \psi_0 \| 
    \leq (1 + 3 c_0 C_{\mathrm{HLS},3} N^{3/2} + 2 c_0^2 C_{\mathrm{HLS},3}^2 N^3) \| \psi_0 \|_{H^2},
\end{equation}
which gives the desired estimate and completes the proof.\end{proof}

Next, we prove~\cref{Key lem: N-body est}. The argument is based on the boundedness of the operator
\begin{equation}\label{Qineq}
    C_{\mathrm{HLS},n} := \left\| \frac{1}{|p_y|} \frac{1}{|y|} \right\|_{L_y^2(\mathbb{R}^n) \to L_y^2(\mathbb{R}^n)} 
    = \left\| \frac{1}{|y|} \frac{1}{|p_y|} \right\|_{L_y^2(\mathbb{R}^n) \to L_y^2(\mathbb{R}^n)} < \infty, \qquad n \geq 3,
\end{equation}
which is established in~\cref{sec: A}. In the case $n = 3$, the constant is explicitly given by
\begin{equation}\label{def CHLS3}
    C_{\mathrm{HLS},3} = 2,
\end{equation}
see~\cite[Theorem 2.5]{Herbst1977}.
\begin{proof}[Proof of~\cref{Key lem: N-body est}] Let \( f \in L^2 \). Using~\cref{eq:N-V_def}, we write
\begin{equation}\label{express: Vp}
    V \frac{1}{|p|} f = \sum_{1 \leq j < k \leq N} \frac{c_{jk}}{|x_j - x_k|} \frac{1}{|p|} f.
\end{equation}
To estimate \( V \frac{1}{|p|} f \), we observe that
\begin{equation}
    \frac{1}{|x_j - x_k|} \frac{1}{|p_j|} = e^{-i x_k \cdot p_j} \frac{1}{|x_j|} \frac{1}{|p_j|} e^{i x_k \cdot p_j},\qquad \forall 1\leq j<k\leq N.
\end{equation}
This equation, together with estimate~\eqref{Qineq}, yields
\begin{equation}\label{eq:1/|x_j_x_k|and_p_j}
    \left\| \frac{1}{|x_j - x_k|} \frac{1}{|p_j|} \right\| \leq \left\| \frac{1}{|x_j|} \frac{1}{|p_j|} \right\| \leq C_{\mathrm{HLS},3}.
\end{equation}
Applying this bound to~\cref{express: Vp}, and using condition~\eqref{con: c0}, we obtain
\begin{equation}\label{line1}
    \left\| V \frac{1}{|p|} f \right\| \leq c_0 C_{\mathrm{HLS},3} \sum_{1 \leq j < k \leq N} \left\| \frac{|p_j|}{|p|} f \right\|.
\end{equation}
Applying the Cauchy--Schwarz inequality to~\eqref{line1}, we get
\begin{equation}\label{line2}
\begin{aligned}
    \left\| V \frac{1}{|p|} f \right\|
    &\leq c_0 C_{\mathrm{HLS},3} \left( \sum_{1 \leq j < k \leq N} 1 \right)^{1/2}
        \left( \sum_{1 \leq j < k  \leq N} \left\| \frac{|p_j|}{|p|} f \right\|^2 \right)^{1/2} \\
    &\leq c_0 C_{\mathrm{HLS},3} N \left( \sum_{1 \leq j < k \leq N} \left\| \frac{|p_j|}{|p|} f \right\|^2 \right)^{1/2}.
\end{aligned}
\end{equation}
We now estimate the second factor. Observe that
\begin{equation}
\begin{aligned}
    \sum_{1 \leq j < k \leq N} \left\| \frac{|p_j|}{|p|} f \right\|^2
    &= \left( f, \sum_{1 \leq j < k \leq N} \frac{|p_j|^2}{|p|^2} f \right)_{L^2} \\
    &= \left( f, \sum_{j=1}^N \frac{(N - j)}{|p|^2} |p_j|^2 f \right)_{L^2} \\
    &\leq N \left( f, \sum_{j=1}^N \frac{|p_j|^2}{|p|^2} f \right)_{L^2} 
    = N \|f\|^2.
\end{aligned}
\end{equation}
Substituting this into~\eqref{line2} yields the desired estimate~\eqref{key est: N-body}, which completes the proof.\end{proof}

\section{$N$-body Trotter Error Estimate}\label{sec:many-body-trotter}
\REV{In this section, we analyze the $N$-body Trotter error by deriving an exact local error operator representation and then carefully estimating each term in the many-body setting. \cref{sec:subsection:error_statements} presents the error representation together with the statements of the main theorems. We present two versions of the error results, one in terms of the initial wavefunction's Sobolev norm, which is what we mainly focus on this paper, we also present an alternative format in terms of the norms of $H\psi(0)$, where $\psi(0)$ is the initial condition. In \cref{sec:subsection:main_thm_pf}, we provide the proofs of the local error bounds, which constitute the main result of this section.
\subsection{Trotter Error Estimates}\label{sec:subsection:error_statements}}
We consider the Trotterization, denoted as $U_1$, given by
\begin{equation}\label{eq:trotter_def}
    U_1(t) : =  e^{-itB} e^{-itA} \approx e^{-i Ht},
\end{equation}
where $A = -\Delta$ and $B = V(x)$ as defined in~\cref{eq:N-V_def}. Its local truncation error admits the following exact error representation. While the proof is elementary, we include it here for completeness, as the form of the representation differs slightly from those typically used for Trotter error analysis in the bounded-operator setting (e.g., \cite[Section 3.1]{ChildsSuTranEtAl2020}, \cite[Lemma 4]{AnFangLin2021}). \REV{In what follows, we derive the error operator $E(t)$, which will later act on initial data in $H^2$, the domain of the Hamiltonian (see~\cref{eq:e_sigma}), yielding the exact representation of the local truncation error. We then analyze this expression to obtain the corresponding error estimate.}

\begin{lemma}[Trotter Local Error Representation]\label{lem:trotter_local_err_rep} Let $E(t)$ denote the difference between the Trotterized evolution $U_1(t)$ and the exact unitary $U(t) = e^{-iHt}$.
\begin{equation}
    E(t) = U_1(t) - U(t) = i\int_0^t ds \, e^{-is B }[e^{-is A}, B]e^{-i(t-s)H}.
\end{equation}
\end{lemma}
\begin{proof}
    The proof follows a straightforward calculation. Specifically, consider \begin{equation}
        \Omega(s) : = e^{-isB}e^{-isA}e^{-i(t-s) H}.
    \end{equation}
   so that $ \Omega(t) = U_1(t) $ and $\Omega(0) = U(t)$. By the fundamental theorem of calculus, one has
   \begin{equation}
       U_1(t) - U(t) =  \int_0^t \,ds \frac{d}{ds}\Omega(s),
   \end{equation}
   where 
   \begin{align}
     \frac{d}{ds}\Omega(s) = & \ e^{-isB}(-iB)e^{-iAs}e^{-i(t-s)H}
          \nonumber
     \\ 
     & \
     +  e^{-isB}e^{-iAs}(-iA)e^{-i(t-s)H}  
 + e^{-isB}e^{-iAs}(iH)e^{-i(t-s)H} \nonumber
     \\
     = & \  e^{-isB}(-iB)e^{-iAs}e^{-i(t-s)H}  + e^{-isB}e^{-iAs}(iB)e^{-i(t-s)H} 
     \nonumber
     \\
     = & \ ie^{-isB}[e^{-iAs},B]e^{-i(t-s)H} .
   \end{align}
\end{proof}
It is worth noting that the above error representation applies generally, independent of the Coulomb interaction setting considered in this work. Thanks to unitarity, the global error of Trotterization is simply upper bounded by the sum of the local errors across all Trotter steps. Another important remark is that for general bounded operators, the commutator $[e^{-isA}, B]$ can be expressed in an integral form, with the integrand bounded by the norm of the commutator $[A, B]$. This aligns with standard approaches in Trotter error analysis, e.g., in~\cite{ChildsSuTranEtAl2020,AnFangLin2021}. To see this, one can apply the fundamental theorem of calculus to $\Gamma(\tau) := e^{-i \tau A} B e^{-i(s-\tau)A}$:
\begin{equation}
[e^{-isA}, B] = \Gamma(s) - \Gamma(0) = \int_0^s \,d \tau  \, \frac{d \Gamma(\tau)}{d\tau}
= -i \int_0^s \, d\tau \, e^{-i\tau A} [A,B] e^{-i(s-\tau)A},
\end{equation}
which is thus bounded above by $\norm{[A,B]}s$ -- this explains why the Trotter error is controlled by the commutator of the summands. However, in our setting, $B = V(x)$ is the Coulomb potential, which is singular and worsens with each derivative. Proceeding further with the commutator form would introduce second-order derivatives acting on the potential $V(x)$, which are difficult to control as discussed in \cref{sec:pf_strategy}.

The global error operator of Trotterization with time step size $t$ over $L$ steps can be expressed as
\begin{equation}
U_1(t)^L - U(t)^L = \sum_{\ell=0}^{L-1} U_1(t)^{L-1-\ell} (U_1(t) - U(t)) U(t)^\ell,
\end{equation}
which acts on the initial wavefunction $\psi(0)$. Taking the $L^2$-norm, we obtain
\begin{align}
\norm{(U_1(t)^L - U(t)^L) \psi(0)} \leq & \sum_{\ell=0}^{L-1} \norm{ (U_1(t) - U(t)) U(t)^\ell \psi(0)}
\\
\leq & \sum_{\ell=0}^{L-1} \norm{\int_0^t ds \, e^{-is B }[e^{-is A}, B]e^{-i(t-s+t \ell)H} \psi(0)}.\label{eq:long_term_error_by_lte}
\end{align}

Hereafter, we focus solely on estimating the local truncation error acting on the initial condition, specifically $\sup\limits_{\sigma\in [0,T]}\|e_\sigma(t)\|$ with $e_\sigma(t), \sigma = t \ell\in [0,T]$ given by
\begin{equation}\label{eq:e_sigma}
    e_\sigma(t):=\int_0^t ds e^{-isV(x)}[e^{-is(-\Delta)}, V(x)]e^{-i(t-s+\sigma)H}\psi(0), \qquad \psi(0)\in H^2.
\end{equation}
Here we recall that $V(x)$ is given in~\cref{eq:N-V_def}. \par

\begin{thm}[local Trotter error]\label{thm Tt} If the condition~\eqref{con: c0} holds, then for the time step size $t \in (0,1]$,
\begin{equation}
   \sup\limits_{\sigma\in [0,T]} \|e_\sigma(t)\| \leq \tilde{C}_N \, t^{\frac{5}{4}} \|\psi(0)\|_{H^2}
\end{equation}
holds true, where $\tilde{C}_N$ is given by
\begin{equation}\label{def CN'}
    \tilde{C}_N := \frac{4}{5} c_0 \tilde{C}_F \left( (N-1)N^{\frac{3}{2}} + (N-1)N^{\frac{1}{2}}(C_N - 1) \right),
\end{equation}
and
\begin{equation}\label{def tCF}
    \tilde{C}_F := \frac{4\sqrt{6}}{3} C_{F1} + 24 C_{F2} C_{\mathrm{HLS},3} + 2,
\end{equation}
with \( C_N \) defined in~\cref{def CN}, and $c_0$, $C_{\mathrm{HLS},3}$, $C_{F1}$, and $C_{F2}$ are all absolute constants  
defined 
in~\cref{con: c0,def-CF1,def-CF2,def CHLS3} with their numerical values provided.
\end{thm}

As noted, $C_{F1}$, $C_{F2}$, and $C_{\mathrm{HLS},3}$ are all absolute constants. In particular,  $C_{\mathrm{HLS},3} = 2$, and $C_{F1}$ and $C_{F2}$ associated with the properties of the smooth cutoff function $F$. While there are many possible choices for the cutoff function, we select a specific one and explicitly compute the corresponding constants as given in~\cref{def-CF1,def-CF2}. We keep $C_{F1}$, $C_{F2}$, and $C_{\mathrm{HLS},3}$ in the theorem instead of substituting in their numerical values, as this form makes it more transparent where each constant originates. The high-level reason the bound remains uniform in $T$ (or $\sigma$) is that the solution's $H^2$ norm is uniformly bounded in time. Our main result (\cref{thm:main_trotter_long}) immediately follows from~\cref{thm Tt} together with \cref{eq:long_term_error_by_lte}. \REV{From the discussion above, it is helpful to emphasize that the prefactor $\tilde C_N$ in \cref{thm:main_trotter_long} is exactly the same as the prefactor in the main result stated in \cref{thm Tt}. In the following remark, we make the universal constant $\tilde C$ in \cref{thm Tt} more explicit.}

\begin{rem}\label{rem:universal_constant}
\REV{We note that the focus of the paper is to show that the error scaling is polynomial in $N$ with a universal constant. Further optimizing the universal constant of $\tilde{C}>0$ is beyond the scope of this work. Nevertheless, we provide a brief remark on its precise value (although this constant can certainly be improved by choosing a better function $F$ and by treating sums of powers in $N$ more carefully). In what follows, we will only use the fact that $N\ge 1$ to crudely bound all lower powers of $N$ by the highest power.
Recall the expression of $C_N$ as in~\cref{def CN}
    \begin{equation} 
    \begin{aligned}
    C_N -1 = & 1 + 3 c_0 C_{\mathrm{HLS},3} N^{3/2} + 2 c_0^2 C_{\mathrm{HLS},3}^2 N^3
    \\
     =& 1 + 6 c_0  N^{3/2} + 8 c_0^2  N^3  \leq (1+6 c_0 + 8 c_0^2) N^3,
    \end{aligned}
\end{equation}
where we used the facts that $C_{\mathrm{HLS},3} = 2$ and $N \geq 1$. From~\cref{def CN'}, we have
\begin{equation}
\begin{aligned}
    \tilde{C}_N & \leq  \frac{4}{5} c_0 \tilde{C}_F \left( N^{\frac{5}{2}} + N^{\frac{3}{2}}(C_N - 1) \right)
    \leq 
    \frac{4}{5} c_0 \tilde{C}_F \left( N^{\frac{5}{2}} + N^{\frac{9}{2}}(1 + 6 c_0  + 8 c_0^2) \right)
    \\
    & \leq 
    \frac{4}{5} c_0 \tilde{C}_F  N^{\frac{9}{2}} (2 + 6 c_0  + 8 c_0^2) 
    ,
\end{aligned}
\end{equation}
where again the inequality uses only $N \geq 1$. Here the constant $\tilde{C}_F$ defined in \cref{def tCF} can be roughly estimated as
\begin{equation}
    \tilde{C}_F = \frac{32\sqrt{6}}{3} e^{26/3} + 48 (1 + 4 e^{\frac{32}{3}})  + 2,
\end{equation}
where we used the loose bounds (see \cref{def-CF1,def-CF2}) of 
\begin{equation}
C_{F1}  \leq 8 e^{26/3}, \quad   C_{F2} \leq 1 + 4 e^{\frac{32}{3}}.
\end{equation}
We note that the estimates in the absolute constants $C_{F1}$ and $C_{F2}$ are very coarse. In principle, they should be optimized over all admissible functions $F$ satisfying \cref{def F}, and the quantities
\begin{equation} 
    C_{F1} := \sup_{\eta \in \mathbb{R}^3} |\eta|^2 \left| F''(|\eta| > 1) \right|,
\quad 
    C_{F2} := \sup_{\eta \in \mathbb{R}^3} \left|\, |\eta| F'(|\eta| > 1) - F(|\eta| > 1) \,\right|,
\end{equation}
should then be computed accurately to obtain their minimal possible values. Here we only provide an upper bound by choosing one particular $F$ (as in~\cref{eq:F_particular}) and estimating a rough upper bound of $C_{F1}$ and $C_{F2}$ by hand, without any attempt at optimization. Further refinement of this universal constant is mathematically interesting, but lies outside the scope of the present work.
}
\end{rem}

\begin{rem}
    \REV{We also remark that an immediate consequence of \cref{thm Tt} (and \cref{thm:main_trotter_long}) is the estimate of the number of Trotter steps $L$, which is given by 
    \begin{equation}
        L  = \mathcal{O}\left (\frac{N^{18} T^{5} \norm{\psi_0}_{H^2}^4}{ \epsilon^4} \right). 
    \end{equation}
}
\end{rem}

\REV{Alternatively, as a byproduct of our proof of \cref{thm Tt}, we can also bound the Trotter local error in terms of the norm of $H\psi_0$, rather than the initial state's Sobolev norm, by invoking \cref{eq:lap_norm_bounded_by_Hpsi}. The two versions of the theorems differ only in the constant dependent on the initial condition and share the same parameter-scaling behavior. This yields the following alternative version of the local error theorem.
\begin{thm}[local Trotter error - Alternative]\label{thm Tt-Hpsi-ver} If the condition~\eqref{con: c0} holds, then for the time step size $t \in (0,1]$,
\begin{equation}
\begin{aligned}
   \sup\limits_{\sigma\in [0,T]} \|e_\sigma(t)\| \leq& \frac{4}{5}c_0 \tilde{C}_F \, t^{\frac{5}{4}} \bigg(
   \left( (N-1)N^{1/2}(1 + 2 c_0N^{3/2}) \right) \|H\psi(0)\| \\
   & \qquad + 
   \left( (N-1) N^{3/2} + (N-1)N^{1/2}\left(2 c_0N^{3/2}  + 4c_0^2 N^3\right) \right) \| \psi(0)\| \bigg)
   \end{aligned}
\end{equation}
holds true, where $\tilde{C}_F$ is the absolute constant defined in \cref{def tCF}.
\end{thm}
}

\subsection{Proof of \cref{thm Tt} \label{sec:subsection:main_thm_pf} and \cref{thm Tt-Hpsi-ver}}
The proof of~\cref{thm Tt} requires the following lemmas (\cref{lem: dDelta,lem: N v,lem:p_j_p_k_counting_in_N}).
Below, we present the statements and proofs of \cref{lem: dDelta,lem:p_j_p_k_counting_in_N}, along with the statement of~\cref{lem: N v}. To ensure a smoother presentation and minimize interruptions to the main argument, the proof of~\cref{lem: N v} is deferred to~\cref{sec:app_n-body}. This is because~\cref{lem: N v} concerns only properties of the three-dimensional Coulomb potential in analogy to the one-body setting and is not a core challenge in the many-body system size counting argument. \REV{After establishing these lemmas, we will prove \cref{thm Tt}. At the end of the section, we also show that the proof of \cref{thm Tt} immediately implies \cref{thm Tt-Hpsi-ver}.}

\REV{As will be made clear later, the following change of variables 
\[
(x_1, x_2) \to (x_1 - x_2, x_1 + x_2)
\]
will come in handy (see~\eqref{eq:change_of_var_x1-x2}); this motivates us to consider the following lemma, where in particular the change of coordinates is 
$
(y, z) \to (y - z,\; y + z).
$}
\begin{lemma}\label{lem: dDelta} For $y = (y_1, y_2, y_3)$ and $z = (z_1, z_2, z_3)$ in $\mathbb{R}^3$, let $p_y := -i\nabla_y$ and $p_z := -i\nabla_z$. Then for all $g(y, z) \in H^2$, we have
\begin{equation}\label{est: g}
    \big\|\, |p_y|\, \partial_{y_j - z_j} g \,\big\|^2 
    \leq \frac{3}{4} \big\|\, |p_y|^2 g \,\big\|^2 + \frac{1}{4} \big\|\, |p_z|^2 g \,\big\|^2,\qquad j=1,2,3.
\end{equation}

\end{lemma}
\begin{proof}
    We note that
    for $j = 1, 2, 3$, we have
\begin{equation}
\begin{aligned}
    \big\|\, |p_y| \, \partial_{y_j - z_j} g \,\big\|^2 
    &= -\big( |p_y| g, \, \partial_{y_j - z_j}^2 |p_y| g \big)_{L^2} \\
    &\leq \frac{1}{2} \big( |p_y| g, \, (-\Delta_y - \Delta_z) |p_y| g \big)_{L^2} \\
    &= \frac{1}{2} \left( \|\, |p_y|^2 g \,\|^2 + \|\, |p_z|\, |p_y| g \,\|^2 \right) \\
    &\leq \frac{3}{4} \|\, |p_y|^2 g \,\|^2 + \frac{1}{4} \|\, |p_z|^2 g \,\|^2, \qquad \forall\, g \in H^3,
\end{aligned}
\end{equation}
where in the first inequality we also used the positivity of $-\Delta$.
By the density of $H^3$ in $H^2$, the same conclusion holds for all $g \in H^2$, yielding the result.
\end{proof}
Let $v\,:\, \mathbb R^3 \setminus \{0\}\to \mathbb R,\, y\mapsto v(y)=\frac{1}{|y|}$. We define 
\begin{equation}\label{def vreg}
    v_{\mathrm{reg}}(y,s):=F\left(\frac{|y|}{s^\beta}>1\right)\frac{1}{|y|}
\end{equation}
and 
\begin{equation}\label{def vsin}
    v_{\mathrm{sin}}(y,s):=F\left(\frac{|y|}{s^\beta}\leq 1\right)\frac{1}{|y|},
\end{equation}
for all $s>0$. We use smooth cutoff functions \( F(\cdot \leq 1) \) and \( F(\cdot > 1) := 1 - F(\cdot \leq 1) \), satisfying:
\begin{equation}\label{def F}
    F(\lambda \leq 1) = 
    \begin{cases}
        1 & \text{for } \lambda \leq \tfrac{1}{2}, \\
        0 & \text{for } \lambda \geq 1.
    \end{cases}
\end{equation}
For example, we can take it as a smooth transition function, namely,
\begin{equation} \label{eq:F_particular}
    F(\lambda \leq 1) = 
\begin{cases}
1 & \lambda \leq 1/2 \\
C_0\int_{\lambda}^1 e^{-\frac{1}{(r-1/2)(1-r)}}dr
& \lambda \in (1/2, 1) \\
0 & \lambda \geq 1
\end{cases}
\end{equation}
with 
\begin{equation}
    C_0:=\dfrac{1}{\int_{\frac{1}{2}}^1 e^{-\frac{1}{(r-1/2)(1-r)}}dr}. 
\end{equation}
It is helpful to note that $F(\lambda > 1) \leq \chi(\lambda > 1/2)$, where $\chi(z\in I)$ denote an indicator function of $z$ on interval $I$.
\begin{lemma}\label{lem: N v} For all $s > 0$ and $y \in \mathbb{R}^3 \setminus \{0\}$, we have
\begin{align}
    \left| [-\Delta v_{\mathrm{reg}}](y, s) \right| &\leq C_{F1} \, \chi\left( |y| > \tfrac{1}{2} s^{\beta} \right) \cdot \frac{1}{|y|^3}, \label{regV-N-body1} \\
    \left| [\partial_{y_j} v_{\mathrm{reg}}](y, s) \right| &\leq C_{F2} \, \chi\left( |y| > \tfrac{1}{2} s^{\beta} \right) \cdot \frac{1}{|y|^2}, \quad y_j := y \cdot e_j,\quad j = 1,2,3, \label{regV-N-body2}
\end{align}
where the constants \( C_{F1} \) and \( C_{F2} \) are defined by
\begin{equation} \label{def-CF1}
    C_{F1} := \sup_{\eta \in \mathbb{R}^3} |\eta|^2 \left| F''(|\eta| > 1) \right|\leq 8 e^{\frac{26}{3}},
\end{equation}
and
\begin{equation} \label{def-CF2}
    C_{F2} := \sup_{\eta \in \mathbb{R}^3} \left|\, |\eta| F'(|\eta| > 1) - F(|\eta| > 1) \,\right|\leq 1 + C_0 \leq 1 + 4 e^{\frac{32}{3}}.
\end{equation}

\end{lemma}
\begin{lemma}\label{lem:p_j_p_k_counting_in_N} We follow the convention $\langle \xi\rangle:=\sqrt{|\xi|^2+1}$ for all $\xi\in \mathbb R^n$. For all \( f \in H^2 \), we have
\begin{equation}\label{sum NT}
    \begin{aligned}
        \sum_{1 \leq j < k \leq N} \left( \| \langle p_j \rangle^2 f \| + \| \langle p_k \rangle^2 f \| \right)
        &\leq (N - 1) N^{3/2} \| f \| + (N - 1) N^{1/2} \| (-\Delta) f \|.
    \end{aligned}
\end{equation}
\end{lemma}
\begin{proof} We note that
\begin{equation}
    \begin{aligned}
        \sum_{1 \leq j < k \leq N} \left( \| \langle p_j \rangle^2 f \| + \| \langle p_k \rangle^2 f \| \right)
        &= \sum_{j=1}^{N-1} \sum_{k=j+1}^N \| \langle p_j \rangle^2 f \| + \sum_{k=2}^{N} \sum_{j=1}^{k-1} \| \langle p_k \rangle^2 f \| \\
        &= (N - 1) \sum_{j=1}^N \| \langle p_j \rangle^2 f \|.
    \end{aligned}
\end{equation}
By the Cauchy–Schwarz inequality, this yields
\begin{equation}
    \begin{aligned}
        \sum_{1 \leq j < k \leq N} \left( \| \langle p_j \rangle^2 f \| + \| \langle p_k \rangle^2 f \| \right)
        &\leq (N - 1) \left( \sum_{j=1}^N 1 \right)^{1/2} \left( \sum_{j=1}^N \| \langle p_j \rangle^2 f \|^2 \right)^{1/2} \\
        &= (N - 1) N^{1/2} \left( f, \sum_{j=1}^N \langle p_j \rangle^4 f \right)^{1/2}_{L^2}.
    \end{aligned}
\end{equation}
This, together with the inequality
\begin{equation}
    \sum_{j=1}^N \langle q_j \rangle^4 \leq \left( \sum_{j=1}^N \langle q_j \rangle^2 \right)^2, \qquad \forall\, q = (q_1, \dots, q_N) \in \mathbb{R}^{3N},
\end{equation}
yields
\begin{equation}
    \begin{aligned}
        \sum_{1 \leq j < k \leq N} \left( \| \langle p_j \rangle^2 f \| + \| \langle p_k \rangle^2 f \| \right)
        &\leq (N - 1) N^{1/2} \| (N + |p|^2) f \| \\
        &\leq (N - 1) N^{3/2} \| f \| + (N - 1) N^{1/2} \| (-\Delta) f \|,
    \end{aligned}
\end{equation}
which completes the proof.
\end{proof}

\begin{proof}[Proof of~\cref{thm Tt}] We write $e_\sigma(t)$ as
\begin{equation}\label{split: et}
    e_\sigma(t) = \sum\limits_{1 \leq j < k \leq N} c_{jk} e_{jk}(t),
\end{equation}
where $e_{jk}(t)\equiv e_{\sigma,jk}(t)$ (we omit the explicit dependence on $\sigma$ for notational simplicity), for $1 \leq j < k \leq N$, are given by
\begin{equation}
    e_{jk}(t) := \int_0^t ds\, e^{-isV(x)} [e^{-is(-\Delta)}, \tfrac{1}{|x_j - x_k|}] e^{-i(t-s+\sigma)H} \psi(0).
\end{equation}
Next, we estimate $\|e_{12}(t)\|$, and the bounds for $\|e_{jk}(t)\|$ (for $1 \leq j < k \leq N$) follow similarly. For $e_{12}(t)$, we estimate $\|[e^{-is(-\Delta)}, \tfrac{1}{|x_1 - x_2|}]\|$. Following the one-body case, decompose the potential $v(x_1 - x_2) := \tfrac{1}{|x_1 - x_2|}$ as
\begin{equation}
    v(x_1 - x_2) = v_{\mathrm{reg}}(x_1 - x_2, s) + v_{\mathrm{sin}}(x_1 - x_2, s),
\end{equation}
where $v_{\mathrm{reg}}$ and $v_{\mathrm{sin}}$ are defined in~\cref{def vreg,def vsin}. We note that
\begin{equation}\label{eq:change_of_var_x1-x2}
\begin{aligned}
    -\Delta =& -\Delta_{x_1} -\Delta_{x_2} + \sum\limits_{j=3}^N \Delta_{x_j}
    \\ 
    = & 2(-\Delta_{x_1 - x_2}) + 2(-\Delta_{x_1 + x_2}) - \sum\limits_{j=3}^N \Delta_{x_j},
    \end{aligned}
\end{equation}
which implies
\begin{equation}
    [-\Delta, v_{\mathrm{reg}}(x_1 - x_2, s)] = 2[-\Delta_{x_1 - x_2}, v_{\mathrm{reg}}(x_1 - x_2, s)].
\end{equation}
Using this and \cref{eq:expand_comm_eA_B}, for 
\begin{equation}
    f = e^{-i(t-s+\sigma)H}\psi(0) \in H^2,
\end{equation}
we compute (with $[-\Delta v_{\mathrm{reg}}](y, s) \equiv -\Delta_y[v_{\mathrm{reg}}(y, s)]$):
\begin{equation}\label{cVrf N}
\begin{aligned}
    &\left\| [e^{-is(-\Delta)}, v_{\mathrm{reg}}(x_1 - x_2, s)] f \right\|
    \leq 2\int_0^s du\, \left\| [-\Delta v_{\mathrm{reg}}](x_1 - x_2, s) e^{i(u-s)(-\Delta)} f \right\| \\
    &\quad + 4 \sum_{j=1}^3 \int_0^s du\, \left\| \partial_{(x_1 - x_2) \cdot e_j} v_{\mathrm{reg}}(x_1 - x_2, s) \partial_{(x_1 - x_2) \cdot e_j} e^{i(u-s)(-\Delta)} f \right\|,
\end{aligned}
\end{equation}
with $\{e_1, e_2, e_3\}$ an orthonormal basis in $\mathbb{R}^3$. Using estimates (\cref{regV-N-body1,regV-N-body2}), we get:
\begin{equation}\label{est Vreg N-body1}
       \| [-\Delta v_{\mathrm{reg}}](x_1-x_2,s) \|_{L^2_{x_1}(\mathbb R^3)} \leq C_{F1}\norm{ \chi\left( |y| > \tfrac{1}{2} s^{\beta} \right) \cdot \frac{1}{|y|^3}}=\frac{4}{3}\sqrt{6\pi}\cdot \frac{C_{F1}}{s^{\frac{3}{2}\beta}},
\end{equation}
\begin{equation}\label{est Vreg N-body2}
   \begin{aligned}
    \left\| |x_1-x_2|\, [\partial_{(x_1-x_2)\cdot e_j} v_{\mathrm{reg}}](x_1-x_2,s) \right\|_{L^\infty_{x_1}(\mathbb R^3)}\leq& C_{F2}\norm{ \frac{\chi(|y|>\frac{1}{2}s^\beta)}{|y|}}_{L^\infty_y(\mathbb R^3)} \\
    = &\frac{2C_{F2}}{s^\beta}.
    \end{aligned}
\end{equation}
Applying estimate 
\begin{equation}\label{eq: <y>}
    \begin{aligned}
        \|\langle y\rangle^{-2} \|_{L^2_y(\mathbb R^3)}=&\left(4\pi\int_0^\infty \frac{|y^2|}{(|y|^2+1)^2}d|y|\right)^{\frac{1}{2}}\\
        \leq & \left(4\pi\int_0^\infty \frac{1}{|y|^2+1}d|y|\right)^{\frac{1}{2}}\\
        =& \sqrt{2}\pi
    \end{aligned}
\end{equation}
and then the Sobolev embedding in the $x_1$ variable
\begin{equation}
\begin{aligned}
    \| e^{i(u-s)(-\Delta)} f \|_{L^\infty_{x_1}(\mathbb{R}^3)} \leq& \frac{1}{(2\pi)^{\frac{3}{2}}}\|\langle y\rangle^{-2}\|_{L^2_y(\mathbb R^3)} \|\langle p_1\rangle^2 f \|_{L^2_{x_1}(\mathbb R^3)}\\
    \leq & \frac{1}{2\sqrt{\pi}}\|\langle p_1\rangle^2 f \|_{L^2_{x_1}(\mathbb R^3)},
\end{aligned}
\end{equation}
together with estimates~\eqref{est Vreg N-body1}, we obtain 
\begin{equation}\label{est: 2ndv}
\begin{aligned}
    &\left\| [-\Delta v_{\mathrm{reg}}](x_1-x_2,s)\, e^{i(u-s)(-\Delta)} f \right\|\\
    \leq & \|   \| [-\Delta v_{\mathrm{reg}}](x_1-x_2,s) \|_{L^2_{x_1}(\mathbb R^3)}  \| e^{i(u-s)(-\Delta)} f \|_{L^\infty_{x_1}(\mathbb{R}^3)}   \|_{L^2_{x_2\cdots x_N}(\mathbb R^{3(N-1)})}\\
    \leq & \frac{2}{3}\sqrt{6}\cdot \frac{C_{F1}}{s^{\frac{3}{2}\beta}}\| \langle p_1\rangle^2 f\|,
\end{aligned}
\end{equation}
where \cref{eq:1/|x_j_x_k|and_p_j} was used.
Applying estimates~\eqref{est Vreg N-body2} and
\begin{equation}
\begin{aligned}
    &\left\| \frac{1}{|x_1-x_2|} \partial_{(x_1-x_2)\cdot e_j} e^{i(u-s)(-\Delta)} f \right\| \\
    =&\left\| \|\frac{1}{|x_1-x_2|} \partial_{(x_1-x_2)\cdot e_j} e^{i(u-s)(-\Delta)} f\|_{L^2_{x_1}(\mathbb R^3)} \right\|_{L^2_{x_2\cdots x_N}(\mathbb R^{3(N-1)})} \\
    \leq &C_{\mathrm{HLS},3} \| |p_1| \partial_{(x_1-x_2)\cdot e_j} f \|,
\end{aligned}
\end{equation}
we obtain
\begin{equation}
\begin{aligned}
    &\left\| [\partial_{(x_1-x_2)\cdot e_j} v_{\mathrm{reg}}](x_1-x_2,s)\, \partial_{(x_1-x_2)\cdot e_j} e^{i(u-s)(-\Delta)} f \right\|\\
    \leq &    \left\| |x_1-x_2|\, [\partial_{(x_1-x_2)\cdot e_j} v_{\mathrm{reg}}](x_1-x_2,s) \right\|_{L^\infty_{x_1}(\mathbb R^3)}    \left\| \frac{1}{|x_1-x_2|} \partial_{(x_1-x_2)\cdot e_j} e^{i(u-s)(-\Delta)} f \right\|\\
    \leq & \frac{2C_{F2}C_{\mathrm{HLS},3}}{s^\beta} \| |p_1|\partial_{(x_1-x_2)\cdot e_j}f\|.
\end{aligned}
\end{equation}
This together with estimate~\eqref{est: g} yields 
\begin{equation}\label{est: 1dv1f}
    \begin{aligned}
        &\left\| [\partial_{(x_1-x_2)\cdot e_j} v_{\mathrm{reg}}](x_1-x_2,s)\, \partial_{(x_1-x_2)\cdot e_j} e^{i(u-s)(-\Delta)} f \right\|\\
    \leq & \frac{2C_{F2}C_{\mathrm{HLS},3}}{s^\beta}  \sqrt{ \frac{3}{4}\||p_1|^2f\|^2+\frac{1}{4}\||p_2|^2f\|^2 }\\
    \leq & \frac{2C_{F2}C_{\mathrm{HLS},3}}{s^\beta}\left(\||p_1|^2f\|+\||p_2|^2f\|\right).
    \end{aligned}
\end{equation}
Estimates~\eqref{est: 2ndv} and~\eqref{est: 1dv1f} together with~\eqref{cVrf N} yield
\begin{equation}\label{est:Vr N}
\begin{aligned}
    \left\| [e^{-is(-\Delta)}, v_{\mathrm{reg}}] f \right\| \leq &\int_0^s du \left( \frac{4\sqrt{6}C_{F1}}{3s^{\frac{3}{2}\beta}} + \frac{24C_{F2}C_{\mathrm{HLS},3}}{s^\beta} \right) \left(\| \langle p_1\rangle^2f \|+\| \langle p_2\rangle^2f \|\right)\\
    \leq& (\frac{4\sqrt{6}}{3}C_{F1}+24C_{F2}C_{\mathrm{HLS},3}) s^{1 - \frac{3}{2}\beta} \left(\| \langle p_1\rangle^2f \|+\| \langle p_2\rangle^2f \|\right)
\end{aligned}
\end{equation}
for $s\in (0,1)$. For the singular part \( v_{\mathrm{sin}}(x_1-x_2,s) \), we use its \( L^2 \)-norm decay to estimate:
\begin{equation}
\begin{aligned}
   & \left\| [e^{-is(-\Delta)}, v_{\mathrm{sin}}(x_1-x_2,s)] f \right\|\\
    \leq& \left\| v_{\mathrm{sin}}(x,s)\, e^{-is(-\Delta)} f \right\| + \left\| e^{-is(-\Delta)} v_{\mathrm{sin}}(x,s)\, f \right\| \\
    \leq& \|\| v_{\mathrm{sin}}(x_1-x_2,s) \|_{L^2_{x_1}(\mathbb R^3)} \cdot \left( \| e^{-is(-\Delta)} f \|_{L^\infty_{x_1}(\mathbb R^3)} + \| f \|_{L^\infty_{x_1}(\mathbb R^3)} \right)\|_{L^2_{x_2\cdots x_N}(\mathbb R^{3(N-1)})},
\end{aligned}
\end{equation}
which together with estimates
\begin{equation}
   \| v_{\mathrm{sin}}(y,s)\|_{L^2_y(\mathbb R^3)}=\left(4\pi\int_0^\infty \left(F(\frac{|y|}{s^\beta}\leq 1)\right)^2 \right)^{\frac{1}{2}}\leq 2\sqrt{\pi}s^{\frac{1}{2}\beta} 
\end{equation}
and 
\begin{equation}
    \begin{aligned}
    \|  f \|_{L^\infty_{x_1}(\mathbb{R}^3)} \leq& \frac{1}{(2\pi)^{\frac{3}{2}}}\|\langle y\rangle^{-2}\|_{L^2_y(\mathbb R^3)} \|\langle p_1\rangle^2 f \|_{L^2_{x_1}(\mathbb R^3)}\\
    (\text{use~\cref{eq: <y>}})\leq & \frac{1}{2\sqrt{\pi}}\|\langle p_1\rangle^2 f \|_{L^2_{x_1}(\mathbb R^3)},
\end{aligned}
\end{equation}
yields 
\begin{equation}
  \left\| [e^{-is(-\Delta)}, v_{\mathrm{sin}}(x_1-x_2,s)] f \right\|  \leq 2\sqrt{\pi} s^{\frac{1}{2}\beta} \times \frac{1}{\sqrt{\pi}}\|\langle p_1\rangle^2 f \|=2s^{\frac{1}{2}\beta}\|\langle p_1\rangle^2 f \|.
\end{equation}
Combining with~\eqref{est:Vr N}, we conclude:
\begin{equation}\label{est:cVf N}
\begin{aligned}
    &\left\| [e^{-is(-\Delta)}, \frac{1}{|x_1-x_2|}] f \right\| \\
    \leq& \left\| [e^{-is(-\Delta)}, v_{\mathrm{reg}}(x_1-x_2,s)] f \right\| + \left\| [e^{-is(-\Delta)}, v_{\mathrm{sin}}(x_1-x_2,s)] f \right\| \\
    \leq& \left( (\frac{4\sqrt{6}}{3}C_{F1}+24C_{F2}C_{\mathrm{HLS},3})s^{1 - \frac{3}{2}\beta} + 2s^{\frac{1}{2} \beta} \right) \left(\| \langle p_1\rangle^2f \|+\| \langle p_2\rangle^2f \|\right).
\end{aligned}
\end{equation}
To optimize the bound, we choose \( \beta = \tfrac{1}{2} \), which equalizes the two powers:
\[
1 - \tfrac{3}{2} \beta = \tfrac{1}{2} \beta \quad \Longrightarrow \quad \beta = \tfrac{1}{2}.
\]
Thus, we obtain the desired bound:
\begin{equation}
    \left\| [e^{-is(-\Delta)}, V] f \right\| \leq \tilde C_{F} s^{\frac{1}{4}}\left( \| \langle p_1\rangle^2f \|+\| \langle p_2\rangle^2f \|\right).
\end{equation}
with $\tilde C_F$ given in~\cref{def tCF}. Therefore, 
\begin{equation}
    \| e_{12}(t)\|\leq \tilde C_F\int_0^t ds \, s^{\frac{1}{4}} \left(\| \langle p_1\rangle^2e^{-i(t-s+\sigma)H}\psi(0) \|+\| \langle p_2\rangle^2e^{-i(t-s+\sigma)H}\psi(0) \|\right).
\end{equation}
Following the same argument, we have 
\begin{equation}
    \| e_{jk}(t)\|\leq \tilde C_F\int_0^t ds \, s^{\frac{1}{4}} \left(\| \langle p_j\rangle^2e^{-i(t-s+\sigma)H}\psi(0) \|+\| \langle p_k\rangle^2e^{-i(t-s+\sigma)H}\psi(0) \|\right),
\end{equation}
for all $1\leq j<k\leq N$. This together with~\cref{split: et} yields 
\begin{equation}\label{est: et}
    \begin{aligned}
        \|e_\sigma(t)\|\leq c_0\tilde C_F\int_0^t ds \, s^{\frac{1}{4}} \sum\limits_{1\leq j<k\leq N} \left(\| \langle p_j\rangle^2e^{-i(t-s+\sigma)H}\psi(0) \|+\| \langle p_k\rangle^2e^{-i(t-s+\sigma)H}\psi(0) \|\right).
    \end{aligned}
\end{equation}
This together with estimates (\cref{thmN est,sum NT}) yields 
\begin{equation}
  \sup\limits_{\sigma\in [0,T]}  \|e_\sigma(t)\|\leq \tilde C_N t^{\frac{5}{4}}\|\psi(0)\|_{H^2},
\end{equation}
where constant $\tilde C_N$ is given in~\cref{def CN'}. This completes the proof.\end{proof}

\REV{\begin{proof}[Proof of \cref{thm Tt-Hpsi-ver}]
   Applying \cref{sum NT} to \cref{est: et} yields
   \begin{equation}
          \begin{aligned}
        \|e_\sigma(t)\|\leq  c_0\tilde C_F\int_0^t ds \, s^{\frac{1}{4}} 
    \bigg( (N - 1) N^{3/2} \| \psi(\tilde t_\sigma)  \| 
+ (N - 1) N^{1/2} \| -\Delta \psi(\tilde t_\sigma) \| \bigg)
        ,
    \end{aligned} 
   \end{equation}
   where $ \tilde t_\sigma= t-s+\sigma$. For the first term, we use conservation of the $L^2$ norm.
For the second term, we apply \cref{eq:lap_norm_bounded_by_Hpsi} to bound
$\| -\Delta \psi(\tilde t_\sigma) \|$.
This gives
   \begin{equation}
       \begin{aligned}
        \|e_\sigma(t)\|\leq & 
        c_0\tilde C_F\int_0^t ds \, s^{\frac{1}{4}} 
    \bigg( (N - 1) N^{3/2} \| \psi(0)  \| 
    \\
&+ (N - 1) N^{1/2} \left((1 + 2 c_0N^{3/2})\norm{H\psi_0} + \left(2 c_0N^{3/2}  + 4c_0^2 N^3\right)\norm{\psi (0)} \right) \bigg)
        ,
    \end{aligned}  
   \end{equation}
   which completes the proof.
\end{proof}}

\section{Conclusion and Remarks}\label{sec:conclusion}
We have established the optimal $1/4$-order convergence rate for first-order Trotterization of many-body quantum systems with Coulomb interactions, explicitly quantifying the polynomial dependence on system size. Our analysis treats the Coulomb potential as an unbounded operator without modification or regularization and applies to all initial states in the domain of the Hamiltonian. The result is independent of any specific spatial discretization, making it, in principle, compatible with both first- and second-quantized circuit constructions across various discretization methods. To the best of our knowledge, this is the first result to achieve this sharp convergence rate--even in the one-body setting--and also the first to establish system-size dependence while treating the Coulomb potential as an unbounded operator without regularization.

As we treat the Coulomb potential as an unbounded operator, our bounds are independent of the spatial discretization scheme (plane waves, Gaussian orbitals, finite elements, real-space grids, etc). In the limit where the basis size $\to \infty$, the observed Trotter scaling must match the continuous-space behavior. 
While many studies of Coulomb potentials fix the spatial basis size and treat it as a given parameter, the resulting cost estimate (number of Trotter steps) depends on that basis size. In practice, however, the basis size is determined by the spatial discretization error, which itself varies with system size $N$ and final time $T$. Consequently, the basis size required to achieve a given target precision should, in principle, depend on both the particle number $N$ and the time $T$. As the basis size increases, the finite-basis asymptotics should converge to our continuous-space predictions, ensuring consistency between finite-resolution simulations and the underlying first-principles theory. \REV{While a direct comparison cannot presently be made due to the lack of a rigorous spatial discretization theory for the unregularized Coulomb potential, we anticipate that future developments will place continuous-space and finite-basis analyses within a complementary framework.}

Moreover, the predicted $1/4$ scaling is observed numerically even with finite spatial discretization. Importantly, these results are not obtained from specially tailored or unphysical initial conditions--they appear even when the initial state is the simple ground-state wavefunction of the hydrogen atom. For example, in practical hydrogen-atom simulations~\cite{BurgarthFacchiHahnJohnssonYuasa2024}, the convergence rate initially follows our $1/4$ prediction as $t$ decreases, with a crossover to first-order convergence emerging only at extremely small $t$. This crossover point shifts to smaller $t$ as the spatial basis size (number of modes) increases, and is expected to shift further for larger system sizes $N$.
The recovery of first-order scaling at very small $t$ is natural: in this regime, the step size is far below the spatial and energy scales of the problem, effectively acting as a strong regularization of the dynamics. However, reaching such a regime requires prohibitively high cost in practice. For practical purposes, the relevant efficiency regime is the initial scaling window, where modest Trotter steps already display the $1/4$ behavior and dominate the cost–accuracy tradeoff.

Although our analysis may appear to require initial conditions in $H^2$, a subspace of $L^2$, this subspace is precisely the domain of the Hamiltonian--that is, the set of all states for which the dynamics is well-defined. In this sense, the result is, in the finite-dimensional analog, an operator-norm-type bound in the worst case, rather than a state-dependent or special-case error bound.

One interesting future direction, currently in progress, is to extend this analysis to higher-order Trotterization.  Such extensions present additional challenges, as $e^{-iVs}$ does not map $H^2$ to itself. In the second-order local error representation, for example,
\begin{equation}
\begin{aligned}
    &   \frac{1}{2}\int_0^t  \int_0^s \int_\tau^{s} e^{ - i(t-s)H} e^{ -is A  /2} 
   e^{-i(s-\alpha) B}  [ [A, B],   B ] e^{-i \alpha B} e^{-i sA/2} \, d\alpha \, d\tau \, ds
    \\
    + & 
     \frac{1}{4}\int_0^t  \int_0^s \int_0^{\tau} e^{-i(t-s) (A+B)} e^{-i(s-\alpha) A /2} 
      [ [A, B],  A ] e^{ -i \alpha A /2} 
     e^{-i s B}  e^{-is A /2} \, d\alpha \, d\tau \, ds,
     \end{aligned}
\end{equation}
for $H = A+B$, it is challenging to avoid such a unitary appearing on the right in exact error representations. 

Furthermore, while our worst-case rate is provably optimal, specific settings--such as restricting to a subspace of the Hamiltonian's domain, low-energy states, or certain observables--may admit faster convergence. Quantifying these improvements is an interesting direction for future research. It would also be valuable to conduct systematic numerical studies of the asymptotic scaling as the system size $N \to \infty$, and to explore situations where the $N$-dependence can be improved, for example, for eigenstates or low-energy subspaces.

Our analysis also paves the way for studying spatial discretization error in many-body Coulomb systems, since in numerical analysis such error is determined by bounds on the solution’s Sobolev norms--which we have obtained as intermediate estimates in our proof. The link between spatial discretization size and the magnitude of spatial derivatives is intuitive: smoother solutions (smaller derivatives) require fewer basis functions to resolve, whereas solutions with larger derivatives demand finer discretization. This makes it possible, as a future direction, to connect spatial discretization requirements directly to the scaling laws established here.

\REV{The main focus of this work is to establish the scaling behavior in $N$ and $t$, and we have not attempted to optimize the universal constant $\tilde{C}$. In future work, it would be interesting both to refine this universal constant and to make the unit dependence more explicit. For electronic systems in atomic units, our Hamiltonian corresponds exactly to the setting considered here. For molecular dynamics applications involving nuclei, however, the nuclear mass $M$ is not equal to $1$ in atomic units. This can be accounted for by treating $\sqrt{M} x$ as the effective coordinate, which modifies the universal constant in our estimate though not affecting the $N$- or $t$-scaling. In this paper, we have not explicitly recorded the resulting dependence on the electron–nucleus mass ratio $m/M \ll 1$ (or equivalently $1/M$ in atomic units). A careful and explicit study of this $1/M$ dependence, together with a refinement of the universal constant appearing in our bounds, is an interesting direction for future work.
}

\REV{While \cite{BurgarthFacchiHahnJohnssonYuasa2024} demonstrated that the $1/4$ rate can be saturated for Coulomb systems via numerical and physical evidence, we remark that no rigorous lower bound is currently known for the unregularized Coulomb potential. Physically, the origin of the bad scaling has been attributed to the high probability of $s$-orbitals encountering the Coulomb singularity, as discussed in \cite{BurgarthFacchiHahnJohnssonYuasa2024}. For short times, our exact error representation allows one to compute the one-step error explicitly and precisely for a given initial state; however, obtaining a rigorous lower bound for long-time evolution appears significantly more challenging. We note that there has been promising progress on lower bounds for Trotterization with bounded Hamiltonians (see, e.g., 
\cite{HahnHartungBurgarthFacchiYuasa2025}), but extending such techniques to unbounded operators remains completely open. Indeed, we expect the bounded and unbounded cases to behave fundamentally differently; for example, the no-fast-forwarding theorem for bosonic systems~\cite{TongAlbertMccleanPreskillSu2022} yields lower bounds of a qualitatively different nature than those for bounded Hamiltonians.
Developing general lower-bound techniques for unbounded Hamiltonians is an interesting and important direction for future work.}

\section*{Acknowledgements}
The authors thank Andrew Baczewski for valuable comments. D.F. acknowledges the support from the U.S. Department of Energy, Office of Science, Accelerated Research in Quantum Computing Centers, Quantum Utility through Advanced Computational Quantum Algorithms, grant no. DE-SC0025572, National Science Foundation via the grant DMS-2347791 and DMS-2438074. X.W. acknowledges the support from Australian Laureate Fellowships, grant FL220100072. A.S. acknowledges the support from National Science Foundation via the grant DMS-2205931. 

\noindent
\textbf{Data Availability}. Data sharing is not applicable to this article, as no data sets were generated or analyzed
during the current study.

\noindent \textbf{Conflict of interest}. On behalf of all authors, the corresponding author states that there is no conflict of interest.

\appendix

\section{Auxiliary Estimates for the One-Body Problem}\label{sec: Aux est one-body}
In this section, we first present the proof of \cref{com lem: one-body}, which represents the only nontrivial component of the analysis and can therefore be effectively regarded as the complete proof of the one-body Trotter error. We then turn to \cref{energy lem: one-body}, whose validity is largely expected to be trivial. Nonetheless, we include a proof without norm equivalence argument to illustrate the idea, as similar techniques will be employed in the many-body case, yielding the system size dependence. As part of \cref{energy lem: one-body}, we first verify that the terms appearing in~\cref{H id} are mathematically well-defined.

\begin{proof}[Proof of~\cref{com lem: one-body}]  To estimate the operator norm of the commutator \( [e^{-is(-\Delta)}, V] \) from \( H^2 \) to \( L^2 \), we decompose the potential \( V \) into a regular (smooth) part and a singular part:
\begin{equation}
    V(x) = V_{\mathrm{reg}}(x,s) + V_{\mathrm{sin}}(x,s),
\end{equation}
where \( V_{\mathrm{reg}} \) and \( V_{\mathrm{sin}} \) are defined in~\cref{def: Vreg,def: Vsin}, respectively. In those definitions, we use smooth cutoff functions \( F(\cdot \leq 1) \) and \( F(\cdot > 1) := 1 - F(\cdot \leq 1) \), where recall that
\begin{equation}
    F(\lambda \leq 1) = 
    \begin{cases}
        1 & \text{for } \lambda \leq \tfrac{1}{2}, \\
        0 & \text{for } \lambda \geq 1.
    \end{cases}
\end{equation}
Take \( f \in H^2 \). To estimate the commutator with the regular part \( V_{\mathrm{reg}}(x,s) \), we compute:
\begin{equation}\label{cVrf}
\begin{aligned}
    \left\| [e^{-is(-\Delta)},V_{\mathrm{reg}}(x,s)]f \right\| 
    &= \left\| (-i) \int_0^s du\, e^{-iu(-\Delta)} [-\Delta, V_{\mathrm{reg}}(x,s)] e^{i(u-s)(-\Delta)} f \right\| \\
    &\leq \int_0^s du\, \left\| [-\Delta V_{\mathrm{reg}}](x,s)\, e^{i(u-s)(-\Delta)} f \right\| \\
    &\quad + 2 \sum_{j=1}^3 \int_0^s du\, \left\| [\partial_{x_j} V_{\mathrm{reg}}](x,s)\, \partial_{x_j} e^{i(u-s)(-\Delta)} f \right\|.
\end{aligned}
\end{equation}
Using the pointwise estimates
\begin{align}
    |[-\Delta V_{\mathrm{reg}}](x,s)| &\lesssim \chi\left(|x| > \tfrac{1}{2} s^{\beta}\right) \cdot \frac{1}{|x|^3}, \label{regV one-body}\\
    |[\partial_{x_j} V_{\mathrm{reg}}](x,s)| &\lesssim \chi\left(|x| > \tfrac{1}{2} s^{\beta}\right) \cdot \frac{1}{|x|^2}, \quad j = 1,2,3,
\end{align}
we obtain the bounds:
\begin{equation}\label{est Vreg one-body}
    \| [-\Delta V_{\mathrm{reg}}](x,s) \| \leq \frac{C}{s^{\frac{3}{2}\beta}}, \qquad
    \left\| |x|\, [\partial_{x_j} V_{\mathrm{reg}}](x,s) \right\|_{L^\infty} \leq \frac{C}{s^\beta}.
\end{equation}
Substituting into~\eqref{cVrf}, and applying the Sobolev embedding
\begin{equation}
    \| e^{i(u-s)(-\Delta)} f \|_{L^\infty(\mathbb{R}^3)} \leq C \| f \|_{H^2},
\end{equation}
together with estimate (\cref{Qineq})
\begin{equation}
    \left\| \frac{1}{|x|} \partial_{x_j} e^{i(u-s)(-\Delta)} f \right\| \leq C_{\mathrm{HLS},3} \| |p| \partial_{x_j} f \| \leq C_{\mathrm{HLS},3} \| f \|_{H^2},
\end{equation}
we conclude:
\begin{equation}\label{est:Vr}
    \left\| [e^{-is(-\Delta)}, V_{\mathrm{reg}}] f \right\| \leq \int_0^s du \left( \frac{C}{s^{\frac{3}{2}\beta}} + \frac{C}{s^\beta} \right) \| f \|_{H^2}
    \leq C s^{1 - \frac{3}{2}\beta} \| f \|_{H^2}\quad \forall s\in (0,1).
\end{equation}
For the singular part \( V_{\mathrm{sin}}(x,s) \), we use its \( L^2 \)-norm decay to estimate:
\begin{equation}
\begin{aligned}
    \left\| [e^{-is(-\Delta)}, V_{\mathrm{sin}}(x,s)] f \right\|
    &\leq \left\| V_{\mathrm{sin}}(x,s)\, e^{-is(-\Delta)} f \right\| + \left\| e^{-is(-\Delta)} V_{\mathrm{sin}}(x,s)\, f \right\| \\
    &\leq \| V_{\mathrm{sin}}(x,s) \| \cdot \left( \| e^{-is(-\Delta)} f \|_{L^\infty} + \| f \|_{L^\infty} \right) \\
    &\leq C s^{\frac{1}{2}\beta} \| f \|_{H^2}.
\end{aligned}
\end{equation}
Combining with~\eqref{est:Vr}, we conclude:
\begin{equation}\label{est:cVf}
\begin{aligned}
    \left\| [e^{-is(-\Delta)}, V] f \right\| 
    &\leq \left\| [e^{-is(-\Delta)}, V_{\mathrm{reg}}(x,s)] f \right\| + \left\| [e^{-is(-\Delta)}, V_{\mathrm{sin}}(x,s)] f \right\| \\
    &\leq C \left( s^{1 - \frac{3}{2}\beta} + s^{\frac{1}{2} \beta} \right) \| f \|_{H^2}.
\end{aligned}
\end{equation}
To optimize the bound, we choose \( \beta = \tfrac{1}{2} \), which equalizes the two powers:
\[
1 - \tfrac{3}{2} \beta = \tfrac{1}{2} \beta \quad \Longrightarrow \quad \beta = \tfrac{1}{2}.
\]
Thus, we obtain the desired bound:
\begin{equation}
    \left\| [e^{-is(-\Delta)}, V] f \right\| \leq C s^{\frac{1}{4}} \| f \|_{H^2}\qquad \forall\, s\in (0,1).
\end{equation}

\end{proof}

\begin{lemma}\label{lem Hid}~\cref{H id} is valid for all $\psi_0\in H^2$ and $t\in \mathbb R$.
\end{lemma}
\begin{proof} By writing  
\[
e^{-itH}\psi_0 = (-\Delta+1)^{-1}(-\Delta+1)e^{-itH}\psi_0,
\]  
and using the identity \( H + 1 = -\Delta + 1 + V \), we obtain  
\begin{equation}\label{id-one-body}
e^{-itH}\psi_0 = (-\Delta+1)^{-1}e^{-itH}(H+1)\psi_0 - (-\Delta+1)^{-1}V e^{-itH}\psi_0.
\end{equation}
By estimate~\eqref{Qineq} and the assumption \( \psi_0 \in H^2 \), we have \( (H+1)\psi_0 \in L^2 \) and \( |p|^{-1} V e^{-itH}\psi_0 \in L^2 \). Indeed,  
\[
\|H\psi_0\| \leq \|(-\Delta)\psi_0\| + \|V|p|^{-1}\| \, \||p|\psi_0\| \leq (1 + |c| C_{\mathrm{HLS},3})\|\psi_0\|_{H^2} < \infty,
\]  
and  
\[
\||p|^{-1}V e^{-itH}\psi_0\| \leq \||p|^{-1}V\| \, \|e^{-itH}\psi_0\| \leq |c| C_{\mathrm{HLS},3} \|\psi_0\| < \infty.
\]
Combining these estimates with~\cref{id-one-body}, we conclude that \( e^{-itH}\psi_0 \in H^1 \). Then, since  
\[
\|V e^{-itH}\psi_0\| \leq \|V|p|^{-1}\| \, \||p| e^{-itH}\psi_0\| \leq |c| C_{\mathrm{HLS},3} \||p| e^{-itH}\psi_0\| < \infty,
\]  
applying~\cref{id-one-body} again yields \( e^{-itH}\psi_0 \in H^2 \), and hence \( (-\Delta) e^{-itH}\psi_0 \in L^2 \). Therefore,~\cref{H id} holds.\end{proof}

\begin{proof}[Proof of~\cref{energy lem: one-body}] \label{pf:lem2_one-body}
By~\cref{H id} and the unitarity of \( e^{-itH} \) on \( L^2 \), we have
\begin{equation}\label{H2dot psit}
    \| (-\Delta) e^{-itH} \psi_0 \| \leq \| H \psi_0 \| + \| V e^{-itH} \psi_0 \|.
\end{equation}
To estimate the second term on the right-hand side, we use the inequality~\eqref{Qineq}, which gives
\begin{equation}\label{V psit}
    \| V e^{-itH} \psi_0 \| \leq \left\| V \frac{1}{|p|} \right\| \cdot \| |p| e^{-itH} \psi_0 \| \leq |c| C_{\mathrm{HLS},3} \| |p| e^{-itH} \psi_0 \|.
\end{equation}
Next, we estimate \( \| |p| e^{-itH} \psi_0 \| \). Applying~\cref{H id} again to the high-frequency component and using~\eqref{Qineq}, we write:
\begin{equation}
    \chi(|p| > 1) |p| e^{-itH} \psi_0 
    = \chi(|p| > 1) \frac{1}{|p|} \left( e^{-itH} H \psi_0 - V e^{-itH} \psi_0 \right).
\end{equation}
Taking $L^2$-norms and applying the triangle inequality:
\begin{equation}
\begin{aligned}
    \| \chi(|p| > 1) |p| e^{-itH} \psi_0 \| 
    &\leq \left\| \chi(|p| > 1) \frac{1}{|p|} \right\| \cdot \| e^{-itH} H \psi_0 \| 
    + \left\| \chi(|p| > 1) \frac{1}{|p|} V \right\| \cdot \| e^{-itH} \psi_0 \| \\
    &\leq \| H \psi_0 \| + |c| C_{\mathrm{HLS},3} \| \psi_0 \|.
\end{aligned}
\end{equation}
For the low-frequency part, we observe:
\begin{equation}
    \| \chi(|p| \leq 1) |p| e^{-itH} \psi_0 \| \leq \| \psi_0 \|.
\end{equation}
Combining the low- and high-frequency bounds, we obtain:
\begin{equation}
    \| |p| e^{-itH} \psi_0 \| \leq \| H \psi_0 \| + (|c| C_{\mathrm{HLS},3} + 1) \| \psi_0 \|.
\end{equation}
Substituting this into~\eqref{V psit} and then into~\eqref{H2dot psit}, we get:
\begin{equation}
\begin{aligned}
    \| e^{-itH} \psi_0 \|_{H^2}
    &\leq \| \psi_0 \| + \| (-\Delta) e^{-itH} \psi_0 \| \\
    &\leq \| \psi_0 \| + \| H \psi_0 \| + |c| C_{\mathrm{HLS},3} \| |p| e^{-itH} \psi_0 \| \\
    &\leq \| \psi_0 \| + \| H \psi_0 \| + |c| C_{\mathrm{HLS},3} \left( \| H \psi_0 \| + (|c| C_{\mathrm{HLS},3} + 1) \| \psi_0 \| \right) \\
    &= (1 + |c| C_{\mathrm{HLS},3} + |c|^2 C_{\mathrm{HLS},3}^2) \| \psi_0 \| 
    + (1 + |c| C_{\mathrm{HLS},3}) \| H \psi_0 \|.
\end{aligned}
\end{equation}
Finally, applying the estimate~\eqref{H psi0}, we obtain the desired bound:
\begin{equation}
\begin{aligned}
    \| e^{-itH} \psi_0 \|_{H^2} 
    &\leq (1 + |c| C_{\mathrm{HLS},3} + |c|^2 C_{\mathrm{HLS},3}^2) \| \psi_0 \| 
    + (1 + |c| C_{\mathrm{HLS},3})^2 \| \psi_0 \|_{H^2} \\
    &\leq (2 + 3 |c| C_{\mathrm{HLS},3} + 2 |c|^2 C_{\mathrm{HLS},3}^2) \| \psi_0 \|_{H^2}.
\end{aligned}
\end{equation}

\end{proof}

\section{Auxiliary Estimates for the $N$-Body Problem}\label{sec:app_n-body}
\begin{proof}[Proof of~\cref{lem: N v}] Using the representation of $-\Delta$ in spherical coordinates and that $v_{\mathrm{reg}}(y,s)$ is radial in the $y$ variable, we obtain 
\begin{equation}
\begin{aligned}
    [-\Delta v_{\mathrm{reg}}](y,s)=&-\frac{\partial^2 v_{\mathrm{reg}}}{\partial |y|^2}-\frac{2}{|y|}\frac{\partial v_{\mathrm{reg}}}{\partial |y|}\\
    =& -\frac{1}{s^{2\beta}}F''(\frac{|y|}{s^\beta}>1)\frac{1}{|y|}+2\frac{1}{s^\beta}F'(\frac{|y|}{s^\beta}>1)\frac{1}{|y|^2}-2F(\frac{|y|}{s^\beta}>1)\frac{1}{|y|^3}\\
    &-\frac{2}{s^\beta}F'(\frac{|y|}{s^\beta}>1)\frac{1}{|y|^2}+\frac{2}{|y|^3}F(\frac{|y|}{s^\beta}>1),
\end{aligned}
\end{equation}
that is, 
\begin{equation}
    [-\Delta v_{\mathrm{reg}}](y,s)=-\frac{|y|^2}{s^{2\beta}}F''(\frac{|y|}{s^\beta}>1)\cdot\frac{1}{|y|^3}.
\end{equation}
This together with~\cref{def F,def-CF1} yields 
\begin{equation}
\begin{aligned}
      |[-\Delta v_{\mathrm{reg}}](y,s)| \leq &\left(\sup\limits_{\eta\in \mathbb R^3}|\eta|^2|F''(|\eta|>1)|\right)\chi\left(|y| > \tfrac{1}{2} s^{\beta}\right) \cdot\frac{1}{|y|^3}\\
      =&C_{F1}\chi\left(|y| > \tfrac{1}{2} s^{\beta}\right) \cdot \frac{1}{|y|^3}.
\end{aligned}
\end{equation}
Next, we compute 
\begin{equation}
    [\partial_{y_j}v_{\mathrm{reg}}](y,s)=\frac{|y|}{s^\beta}F'(\frac{|y|}{s^\beta}>1)\cdot\frac{y_j}{|y|^3}-F(\frac{|y|}{s^\beta}>1)\cdot\frac{y_j}{|y|^3}.
\end{equation}
This together with~\cref{def F,def-CF2} yields 
\begin{equation}
\begin{aligned}
    | [\partial_{y_j}v_{\mathrm{reg}}](y,s)|\leq &\left(\sup\limits_{\eta\in \mathbb R^3} \left| |\eta|F'(|\eta|>1)-F(|\eta|>1)\right|\right) \chi(|y|>\frac{1}{2}s^{\beta})\cdot \frac{1}{|y|^2}\\
    =&C_{F2}\chi(|y|>\frac{1}{2}s^{\beta})\cdot\frac{1}{|y|^2}
\end{aligned}
\end{equation}
for all $y\in \mathbb R^3\setminus\{0\}$ and $j=1,2,3.$ We now estimate \( C_{F1} \) and \( C_{F2} \). Since the support of \( F'(|\eta| > 1) \) and \( F''(|\eta| > 1) \) is contained in the interval \( [\frac{1}{2}, 1] \), we have
\begin{equation}
    C_{F1} = \sup_{\eta \in \mathbb{R}^3} |\eta|^2 \left| F''(|\eta| > 1) \right| \leq \sup_{\eta \in \mathbb{R}^3} \left| F''(|\eta| > 1) \right|,
\end{equation}
and
\begin{equation}\label{est: CF2}
    C_{F2} := \sup_{\eta \in \mathbb{R}^3} \left|\, |\eta| F'(|\eta| > 1) - F(|\eta| > 1) \,\right| \leq 1 + \sup_{\eta \in \mathbb{R}^3} \left| F'(|\eta| > 1) \right|.
\end{equation}
We compute, for \( \lambda \in \left(\frac{1}{2}, 1\right) \),
\begin{equation}\label{com: F'}
    F'(|\eta| > 1) = -F'(|\eta| \leq 1) = C_0 e^{-\frac{1}{(\lambda - 1/2)(1 - \lambda)}},
\end{equation}
and
\begin{equation}\label{eq: F''}
    F''(|\eta| > 1) = -F''(|\eta| \leq 1) = C_0 \frac{d}{d\lambda} \left[-\frac{1}{(\lambda - 1/2)(1 - \lambda)}\right] e^{-\frac{1}{(\lambda - 1/2)(1 - \lambda)}}.
\end{equation}
Since for all \( r \in \left[\frac{5}{8}, \frac{7}{8}\right] \),
\begin{equation}
    \frac{1}{(r - 1/2)(1 - r)} = 2\left( \frac{1}{r - \frac{1}{2}} + \frac{1}{1 - r} \right) \geq 2\left( \frac{1}{\frac{7}{8} - \frac{1}{2}} + \frac{1}{1 - \frac{5}{8}} \right) = \frac{32}{3},
\end{equation}
we obtain
\begin{equation}\label{est: C0}
    C_0 \leq \frac{1}{\int_{\frac{5}{8}}^{\frac{7}{8}} e^{-\frac{1}{(r - 1/2)(1 - r)}} \, dr} \leq \frac{1}{\int_{\frac{5}{8}}^{\frac{7}{8}} e^{-\frac{32}{3}} \, dr} = 4 e^{\frac{32}{3}}.
\end{equation}
Combining this with~\eqref{est: CF2} and~\eqref{com: F'}, we get
\begin{equation}
    C_{F2} \leq 1 + C_0 \sup_{\lambda \in \left[\frac{1}{2}, 1\right]} e^{-\frac{1}{(\lambda - 1/2)(1 - \lambda)}} \leq 1 + C_0 \leq 1 + 4 e^{\frac{32}{3}}.
\end{equation}
Next, for all \( \lambda \in \left(\frac{1}{2}, 1\right) \), we compute
\begin{equation}\label{eq: dLambda}
\begin{aligned}
    \left| \frac{d}{d\lambda} \left[ -\frac{1}{(\lambda - 1/2)(1 - \lambda)} \right] \right|
    &= \left| -2 \frac{d}{d\lambda} \left[ \frac{1}{\lambda - \frac{1}{2}} + \frac{1}{1 - \lambda} \right] \right| \\
    &= \left| \frac{2(1 - \lambda)^2 - 2(\lambda - \frac{1}{2})^2}{(\lambda - \frac{1}{2})^2 (1 - \lambda)^2} \right| \\
    &\leq \frac{1}{2(\lambda - \frac{1}{2})^2 (1 - \lambda)^2}.
\end{aligned}
\end{equation}
Using~\cref{eq: F''}, the estimates~(\cref{est: C0,eq: dLambda}) and the bound
\begin{equation}
    \sup_{\beta \geq 0} \beta^2 e^{-\beta} = \left. \beta^2 e^{-\beta} \right|_{\beta = 2} = 4 e^{-2},
\end{equation}
we obtain
\begin{equation}
    C_{F1} \leq |F''(|\eta| > 1)| \leq 4 e^{\frac{32}{3}} \cdot \frac{1}{2} \sup_{\beta \geq 0} \beta^2 e^{-\beta} = 8 e^{\frac{26}{3}}.
\end{equation}

\end{proof}
\section{Proof of the estimate~\eqref{Qineq}}\label{sec: A}

For completeness, we provide an elementary proof of \cref{Qineq} for all $n \geq 3$. However, we note that while our proof applies to any $n \geq 3$, it does not yield the sharp constant in the case $n = 3$. A more precise bound and proof for $n = 3$ can be found in~\cite{Herbst1977}, where it is shown that $C_{\mathrm{HLS},3} =  2^{-1} \cdot \frac{ \Gamma\left( \frac{1}{4} \right) }{ \Gamma\left( \frac{5}{4} \right) }  =  2$.

\begin{proof}[Proof of~\eqref{Qineq}] 
Since
\begin{equation}\label{app est1}
    \left\|\chi(|y|\geq 1) \frac{1}{|y|}\frac{1}{|p_y|}\chi(|p_y|\geq 1)\right\|_{L^2_y(\mathbb{R}^n) \to L^2_y(\mathbb{R}^n)} \leq 1, \qquad \forall n \geq 1,
\end{equation}
and
\begin{equation}\label{app est2}
    \left\|\chi(|y|< 1) \frac{1}{|y|}\frac{1}{|p_y|}\chi(|p_y|< 1)\right\|_{L^2_y(\mathbb{R}^n) \to L^2_y(\mathbb{R}^n)} \leq C_n, \qquad \forall n \geq 3,
\end{equation}
where $C_n$ is given by, with $\Gamma$ being the Gamma function, 
\begin{equation}
 C_n=   \|\frac{\chi(|y|<1)}{|y|}\|_{L^2_y(\mathbb R^n)}^2=\int_{S^{n-1}}\left(\int_0^1 |y|^{n-3}d|y|\right)d\sigma(y)=\frac{2\pi^{n/2}}{(n-2)\Gamma(n/2)},
\end{equation}
and since by duality,
\begin{equation}\label{dual1}
\begin{aligned}
    &\left\|\chi(|y|< 1) \frac{1}{|y|}\frac{1}{|p_y|}\chi(|p_y|\geq 1)\right\|_{L^2_y(\mathbb{R}^n) \to L^2_y(\mathbb{R}^n)} \\
    =&  \left\|\chi(|p_y|\geq 1)\frac{1}{|p_y|} \frac{1}{|y|}\chi(|y|< 1)\right\|_{L^2_y(\mathbb{R}^n) \to L^2_y(\mathbb{R}^n)},
    \end{aligned}
\end{equation}
it suffices to prove that
\begin{equation}
    \left\|\chi(|y|< 1) \frac{1}{|y|}\frac{1}{|p_y|}\chi(|p_y|\geq 1)\right\|_{L^2_y(\mathbb{R}^n) \to L^2_y(\mathbb{R}^n)} \leq C_n, \qquad \forall n \geq 3,
\end{equation}
for some constant $ C_n > 0 $ depending on $ n $. For this, we let $\chi(z\in I)$ denote an indicator function of $z$ on interval $I$ and let
\begin{equation}
    \chi_j\, :\, L^2(\mathbb R^n)\to L^2(\mathbb R^n),\quad f(y)\mapsto \chi(|y|\in [2^{-j-1}, 2^{-j}))f(y),\qquad j\in \mathbb{Z}
\end{equation}
and
\begin{equation}
    \hat\chi_k\, :\, L^2(\mathbb R^n)\to L^2(\mathbb R^n),\quad f(y)\mapsto \chi(|p_y|\in [2^{k}, 2^{k+1}))f(y),\qquad k\in \mathbb{Z}.
\end{equation}
We take $f, g\in C_0^\infty(\mathbb R^n)$ and then decompose 
\begin{equation}\label{eq: Qfg}
    Q_{f,g}:=(f, \chi(|y|<1)  \frac{1}{|y|}\frac{1}{|p_y|}\chi(|p_y|\geq 1) g)_{L^2_y(\mathbb R^n)}
\end{equation}
into several pieces:
\begin{equation}
     Q_{f,g}=\sum\limits_{j,k\in \mathbb N} Q_{f,g,j,k}\label{decom Qfg}
\end{equation}
where
\begin{equation}
    Q_{f,g,j,k}:=(f, \chi_j  \frac{1}{|y|}\frac{1}{|p_y|}\hat\chi_k g)_{L^2_y(\mathbb R^n)},\qquad j,k\in \mathbb N.
\end{equation}
We note that for $k\geq j$,
\begin{equation}
    \begin{aligned}
        |Q_{f,g,j,k}|\leq & \|\chi_jf\| \|\hat \chi_k g\| \| \chi_j \frac{1}{|y|}\|_{L^2_y(\mathbb R^n)\to L^2_y(\mathbb R^n)}\|  \frac{1}{|p_y|}\hat\chi_k\|_{L^2_y(\mathbb R^n)\to L^2_y(\mathbb R^n)}\\
        \leq &\frac{1}{2^{k-j-1}}\|\chi_jf\| \|\hat \chi_k g\| 
    \end{aligned}
\end{equation}
and for $k<j$, $n\geq 3$,
\begin{equation}
    \begin{aligned}
        |Q_{f,g,j,k}|\leq & \|\chi_jf\| \| \frac{\chi(|y|\in [2^{-j-1},2^{-j}))}{|y|}\|_{L_y^2(\mathbb R^n)} \|\frac{1}{|p_y|}\hat \chi_k g\|_{L^\infty_y(\mathbb R^n)}  \\
        \leq&   \|\chi_jf\| \| \frac{\chi(|y|\in [2^{-j-1},2^{-j}))}{|y|}\|_{L_y^2(\mathbb R^n)} \| \frac{\chi(|p_y|\in [2^{k},2^{k+1}))}{|p_y|}\|_{L_{p_y}^2(\mathbb R^n)}\|\hat \chi_k g\| \\
        \leq &\frac{4\pi^n}{2^{(\frac{n}{2}-1)(j-k-1)}\left(\Gamma(\frac{n}{2}+1)\right)^2}\|\chi_jf\| \|\hat \chi_k g\|, 
    \end{aligned}
\end{equation}
where we used 
\begin{equation}
    \begin{aligned}
        & \| \frac{\chi(|y|\in [2^{-j-1},2^{-j}))}{|y|}\|_{L_y^2(\mathbb R^n)} \| \frac{\chi(|p_y|\in [2^{k},2^{k+1}))}{|p_y|}\|_{L_{p_y}^2(\mathbb R^n)}\\
        =& \left(2^{-j(n/2-1)}\| \frac{\chi(|y|\in [2^{-1},1))}{|y|}\|_{L_y^2(\mathbb R^n)}\right) \left(2^{(k+1)(n/2-1)}\| \frac{\chi(|p_y|\in [2^{-1},1))}{|p_y|}\|_{L_{p_y}^2(\mathbb R^n)}\right)\\
        =& \frac{1}{2^{(\frac{n}{2}-1)(j-k-1)}}\| \frac{\chi(|y|\in [2^{-1},1))}{|y|}\|_{L_y^2(\mathbb R^n)}^2
    \end{aligned}
\end{equation}
and 
\begin{equation}
    \begin{aligned}
         \| \frac{\chi(|y|\in [2^{-1},1))}{|y|}\|_{L_y^2(\mathbb R^n)}^2\leq& 2^2\| \chi(|y|\in [2^{-1},1))\|_{L_y^2(\mathbb R^n)}^2\\
         \leq & 4\| \chi(|y|\in [0,1))\|_{L_y^2(\mathbb R^n)}^2\\
         =& \frac{4\pi^n}{\left(\Gamma(\frac{n}{2}+1)\right)^2}.
    \end{aligned}
\end{equation}
These estimates together with~\cref{decom Qfg} yield, with $C:=\frac{4\pi^n}{\left(\Gamma(\frac{n}{2}+1)\right)^2}$, 
\begin{equation}
    \begin{aligned}
        |Q_{f,g}|\leq & \sum\limits_{j,k\in \mathbb N} |Q_{f,g,j,k}|\\
        \leq & \sum\limits_{j,k\in \mathbb N} (\frac{\chi(k\geq j)}{2^{k-j-1}}+\frac{C\chi(k<j)}{2^{(\frac{n}{2}-1)(j-k-1)}})\|\chi_jf\| \|\hat \chi_k g\|\\
        \leq &  \sum\limits_{j\in \mathbb N, l\in \mathbb Z} (\frac{\chi(l\geq 0)}{2^{l-1}}+\frac{C\chi(l>0)}{2^{(\frac{n}{2}-1)(l-1)}})\|\chi_jf\| \|\hat \chi_{j-l} g\|\\
        \leq &\sum\limits_{ l\in \mathbb Z} (\frac{\chi(l\geq 0)}{2^{l-1}}+\frac{C\chi(l>0)}{2^{(\frac{n}{2}-1)(l-1)}}) \|f\|\|g\|\\
        =& (4+\frac{4\pi^n}{(2^{\frac{n}{2}-1}-1)\left(\Gamma(\frac{n}{2}+1)\right)^2}) \|f\|\|g\|,
    \end{aligned}
\end{equation}
where we used 
\begin{equation}
    \begin{aligned}
        \sum\limits_{j\in \mathbb N} \|\chi_jf\| \|\hat \chi_{j-l} g\|\leq & \left( \sum\limits_{j\in \mathbb Z} \|\chi_jf\|^2\right)^{\frac{1}{2}}\left( \sum\limits_{j\in \mathbb Z} \|\hat\chi_{j-l}g\|^2\right)^{\frac{1}{2}}\\
        =& \|f\|\|g\|.
    \end{aligned}
\end{equation}
This together with relations~(\cref{dual1,eq: Qfg}) and estimates~\eqref{app est1} and~\eqref{app est2} yields~\eqref{Qineq} with $C_{HLS,n}$ satisfying
\begin{equation}\label{CHLSn}
    C_{HLS,n} \leq 9+\frac{2\pi^{n/2}}{(n-2)\Gamma(n/2)}+\frac{8\pi^n}{(2^{\frac{n}{2}-1}-1)\left( \Gamma(\frac{n}{2}+1)\right)^2}.
\end{equation}

\end{proof}

\bigskip

 \bibliographystyle{unsrt}
 \bibliography{bib}
\end{document}